\documentclass[letterpaper,11pt]{article}

\usepackage{jheppub}
\usepackage{multirow}

\usepackage{subcaption}
\usepackage[countmax]{subfloat}
\usepackage{amssymb}
\usepackage{amsmath}
\usepackage{color}
\usepackage{graphicx}
\usepackage{verbatim}
\usepackage{amsthm}
\usepackage{slashed}
\usepackage{hyperref}
\usepackage{makecell}

\preprint{}

\allowdisplaybreaks

\title{Hadronic decays of Higgs boson at NNLO matched with parton shower}

\author[a]{YaLu Hu}
\emailAdd{018072910016@sjtu.edu.cn}

\author[a]{ChuanLe Sun}
\emailAdd{chlsun60@sjtu.edu.cn}

\author[a]{XiaoMin Shen}
\emailAdd{xmshen137@sjtu.edu.cn}

\author[a,b,c]{Jun Gao}
\emailAdd{jung49@sjtu.edu.cn}

\affiliation[a]{INPAC, Shanghai Key Laboratory for Particle Physics and Cosmology,
School of Physics and Astronomy, Shanghai Jiao Tong University, Shanghai 200240, China}
\affiliation[b]{Key Laboratory for Particle Astrophysics and Cosmology, Shanghai 200240, China
}
\affiliation[c]{Center for High Energy Physics, Peking University, Beijing 100871, China}

\abstract{
We present predictions for hadronic decays of the Higgs boson 
at next-to-next-to-leading order (NNLO) in QCD matched with parton shower based
on the POWHEG framework.
Those include decays into bottom quarks with full bottom-quark mass dependence,
light quarks, and gluons in the heavy top quark effective theory.
Our calculations describe exclusive decays of the Higgs boson
with leading logarithmic accuracy in the Sudakov region and next-to-leading
order (NLO)
accuracy matched with parton shower in the three-jet region, with normalizations fixed to the
partial width at NNLO.
We estimated remaining perturbative uncertainties taking typical
event shape variables as an example and demonstrated the need of future
improvements on both parton shower and matrix element calculations.
The calculations can be used immediately in
evaluations of the physics performances of detector designs for future Higgs
factories.
}

\begin{document}

\maketitle

\newpage

\section{Introduction}

The successful operation of the LHC and the ATLAS and CMS experiments
have led to the discovery of the Higgs boson and completion of the
standard model (SM) of particle physics~\cite{Aad:2012tfa, Chatrchyan:2012xdj}.
Precision test on properties of the Higgs boson including all its couplings
with standard model particles becomes one primary task of particle physics
at the high energy frontier.
There have been proposals for possible Higgs factories to measure the Higgs
properties with higher accuracy and to probe possible new physics beyond the
standard model.
That includes electron-positron colliders like ILC~\cite{Behnke:2013xla},
CEPC~\cite{CEPCStudyGroup:2018ghi}, CLIC~\cite{Lebrun:2012hj} and
FCC-$ee$~\cite{Gomez-Ceballos:2013zzn}, or even a muon collider with
the technology advances~\cite{1901.06150}.
In the SM the Higgs boson decays dominantly to hadronic final states which
are hard to access at hadron colliders due to huge QCD backgrounds.
That includes decay channels of a pair of bottom quarks or charm quarks,
a pair of gluons via heavy-quark loops, and four quarks via electroweak gauge
bosons, adding to a total decay branching fraction of about 80\%~\cite{1307.1347}.
The small backgrounds and low hadron multiplicities at lepton collisions
make them an ideal environment to study hadronic decays
of the Higgs boson.
For a Higgs factory like CEPC, with the designed total luminosity we expect an
experimental precision of about one percent for the hadronic decay channels~\cite{1810.09037}. 
On the theory side the partial widths have already been calculated to a very high
accuracy with the intrinsic uncertainties being at percent level or better~\cite{1906.05379}.
The partial width for $H \to b\bar{b}$
is known up to the next-to-next-to-next-to-next-to-leading order (N$^4$LO) in QCD, in
the limit where the mass of the bottom quark is neglected~\cite{Davies:2017xsp}.
The partial width for $H \to gg$ has been calculated to the N$^3$LO~\cite{Baikov:2006ch}
and N$^4$LO~\cite{Herzog:2017dtz} in QCD 
in the heavy top-quark limit.
We refer the readers to~\cite{Denner:2011mq, Spira:2016ztx} for a complete
list of relevant calculations.
On the other hand modelings on hadronic decays at the exclusive level are equally
important.
For instance the full kinematic and flavor information in hadronic decays
will be crucial for correcting for experimental efficiencies~\cite{1905.12903}, especially in the case
where the accompanied $Z$ boson also decays hadronically~\cite{1810.09037,1812.09478}.
Besides, there are also strong motivations for study of 
exclusive hadronic decays to test Yukawa coupling of light quarks~\cite{1608.01746}, and
to search for possible CP violations in Yukawa couplings~\cite{2009.02000} or
exotic states beyond the SM~\cite{0909.1521,1312.4992,Liu:2016zki,Liu:2016ahc,1905.04865}.
The fully differential cross sections
for $H \to b\bar{b}$ have been calculated to next-to-next-to-leading order (NNLO)
in~\cite{Anastasiou:2011qx, DelDuca:2015zqa}
and N$^3$LO in~\cite{Mondini:2019gid} for massless bottom quarks, and to
NNLO in~\cite{Bernreuther:2018ynm,1907.05398,1911.11524,2007.15015}
with massive bottom quarks.
There have been recent predictions
for the hadronic event shapes
in the decay of the Higgs boson at next-to-leading order (NLO)~\cite{Li:2018qiy,1903.07277,Gao:2020vyx}
and approximated NNLO~\cite{1901.02253}.
There also exist a NNLO calculation for the Higgs boson decaying into
a pair of bottom quarks plus an additional jet for
massless bottom quarks~\cite{Mondini:2019vub}.
To achieve a fully exclusive description at hadron level, the fixed-order calculations
need to be further matched with parton shower for example as implemented in the
general purpose MC event generators~\cite{0803.0883,Sjostrand:2014zea,0811.4622}.
Matching schemes at NLO like MC@NLO~\cite{hep-ph/0204244} and POWHEG~\cite{0709.2092} have been
widely used.
There have also been developments on matching NNLO matrix elements with parton shower
for specific processes~\cite{1212.4504,1309.0017,1311.0286,1405.3607,1407.3773,1407.2940,1508.01475,1603.01620,1805.09857,1908.06987,1909.02026,2006.04133,2010.10478,2010.10498}.
Very recently there have been two implementations towards matching hadronic decays of
the Higgs boson at NNLO with parton shower.
Ref.~\cite{1912.09982} presents the matched results for the Higgs boson decaying into
bottom quarks with the MiNLO method~\cite{1206.3572} within POWHEG framework~\cite{1002.2581}
neglecting masses of the bottom quark in the matrix elements.
As noticed there neglecting masses of the bottom quark leads to certain mismatches
with parton shower.
Ref.~\cite{2009.13533} calculates decays to massless bottom quarks as well as to
gluons within the GENEVA framework~\cite{1211.7049}.
In this work we have presented a matched calculation for the Higgs boson decaying into
bottom quarks with full bottom-quark mass dependence within the POWHEG framework.
Besides the mass effects our calculation also differs with Ref.~\cite{1912.09982} in
the way of merging samples with different multiplicities.
In addition we present matched results for the Higgs boson decaying into light quarks and
gluons with similar approaches.
The rest of our paper is organized as follows.
In Sec.~\ref{sec:fra}, we present the theoretical framework used
for our fixed-order calculation as well as the matched calculation.
Sec.~\ref{sec:num} provides numerical results on the partial width
and distributions at various levels.
Finally our summary and conclusions are presented in Sec.~\ref{sec:con}.

\section{Framework}\label{sec:fra}

\subsection{Effective Lagrangian}

Since the top quarks only appear as internal states, one can adopt an effective
theory by integrating out the top quark, the interactions can be expressed as
\begin{equation}\label{eq:leff}
\mathcal{L}_{eff}=\mathcal{L}_{QCD}-\frac{H}{v}C_1G_{\mu\nu}^aG^{\mu\nu,a}
-\frac{H}{v}C_2m_b\bar \psi_{b}\psi_{b},
\end{equation}
to the leading power of inverse of the top-quark mass.
We have neglected the masses of light quarks except for the bottom quark.
We use the on-shell scheme for renormalization of the gluon field, the bottom quark field and
mass, except for the $\overline{\rm MS}$ running mass in the Yukawa
coupling.
The renormalization of QCD coupling is carried out with
a $\overline {\rm MS}$ scheme with $n_l=5$ light flavors.
Moreover, it further requires renormalization of the Wilson coefficients
including mixing effects,
\begin{equation}
C_1^0=Z_{11}C_1, \quad \quad C_2^0=Z_{21}C_1+C_2,
\end{equation}
with the one-loop renormalization constants in the $\overline{\rm MS}$ scheme given by~\cite{hep-ph/9708255},
\begin{eqnarray}
Z_{11}&=&1-\frac{\alpha_S^{(n_l)}(\mu)}{4\pi}S_{\epsilon}\frac{2\beta_0^{(n_l)}}{\epsilon},\nonumber\\
Z_{21}&=&-\frac{\alpha_S^{(n_l)}(\mu)}{4\pi}S_{\epsilon}\frac{12C_F}{\epsilon},
\end{eqnarray}
with $S_{\epsilon}=(4\pi)^{\epsilon}/\Gamma(1-\epsilon)$, and
$\beta_0^{(n_l)}=(11C_A-4T_Rn_l)/6$.
The renormalized Wilson coefficients $C_1$ and 
$C_2$~\cite{Kataev:1981gr,Inami:1982xt,Dawson:1990zj,Djouadi:1991tka,Kataev:1993be,hep-ph/9405325,Spira:1995rr,hep-ph/9708255} carry
further logarithmic dependence on the top-quark mass.
The results at NNLO are given by
\begin{eqnarray}\label{eq:wil}
  C_1&=&-\frac{\alpha_S^{(n_l)}(\mu)}{12\pi}\Big\{1+
\frac{\alpha_S^{(n_l)}(\mu)}{4\pi}
  11+\left(\frac{\alpha_S^{(n_l)}(\mu)}{4\pi}\right)^2\Big[L_t(19+\frac{16}{3}n_l)\nonumber\\
  &&+\frac{2777}{18}-\frac{67}{6}n_l\Big]+\mathcal{O}(\alpha_S^3)
\Big\}, \nonumber \\
  C_2&=&1+\left(\frac{\alpha_S^{(n_l)}(\mu)}{4\pi}\right)^2\Big[{40\over 9}
  -{16\over 3}L_t\Big]+\mathcal{O}(\alpha_S^3),
\end{eqnarray}
with $L_t=\ln(\mu^2/m_t^2)$.
For hadronic decays of the Higgs boson, one usually distinguishes between
decaying into bottom quarks and into gluons. 
Such a separation is only apparent at leading order.
In the following when we refer to the bottom-quark channel or gluon channel
we mean decays initiated by the coupling $C_2$ or $C_1$
respectively.
We have not included the cross terms of $C_1$ and $C_2$ which are formally
of order $\alpha_S^2$ comparing with the decay at leading order~\cite{Davies:2017xsp,Herzog:2017dtz}.
In the calculation of decaying into bottom quarks we keep full
dependence on the bottom-quark mass in the matrix elements.
For the calculation of the gluon channel we set the bottom-quark mass to zero.
Alternatively for decaying into bottom quarks one may also set the bottom-quark
mass in the matrix elements to zero thus neglect associated power corrections.
We refer such a calculation as for decays to massless or light quarks in the sense
that it can be applied directly to light-quark decay channels induced by
various new physics beyond the standard model.

\subsection{QCD factorization and fixed-order calculation}

The starting point of our calculation is to reproduce the partial width
at NNLO in QCD.
We rely on the QCD factorization formulas as derived in either heavy-quark
effective theory (HQEF) or soft-collinear effective theory (SCET).
For instance, in the case where the Higgs boson decays into massive bottom quarks,
we choose a principal variable that is the total radiation-energy, $E_X$,
in the rest frame of the Higgs boson.
In the soft limit, $E_X\ll m_b$, the partial width can be factorized as~\cite{1408.5150,1410.3165},
\begin{equation}
\frac{1}{\Gamma_0}\frac{d\Gamma}{dE_X}=H(Q^2, m_b^2,\mu)\int
\mathrm{d}kS(k,\mu)
\delta(E_X-k),
\end{equation}
where $\mu$ is the renormalization scale, $m_b$ is the bottom quark mass,
and $Q=m_H$ is the typical hard scale of the process.
The soft function for radiation off massive quarks has been calculated
to two-loop level~\cite{1408.5134}.
The hard function at NNLO can be extracted from the two-loop form factors
calculated in Refs.~\cite{hep-ph/0508254} and~\cite{1712.09889}.
We have used results of both groups and find full agreement on the
hard function.
For completeness we include the analytic expressions of the hard function and
the soft function at one-loop level in Appendix~\ref{sec:appb}.
The two-loop results are lengthy and are not shown for simplicity.
In the cases where the Higgs boson decaying into massless quarks or gluons we use $\tau$
variable, $\equiv 1-\mathcal{T}$, with the event shape variable thrust defined as
\begin{equation}
  \mathcal{T}=\max_{ {n}}\frac{\sum_{i}\vert {P}_i\cdot {{n}}
  \vert}{\sum_{i}\vert {P}_i \vert},
\end{equation}
and ${P}_i$ are three-momentum of final state partons, the unit vector 
${n}$ is selected to maximize the projections.
Based on SCET, a factorization formula is derived for $\tau$ in the two-jet
region~\cite{0803.0342},
\begin{equation}
\frac{1}{\Gamma_0}\frac{d\Gamma}{d\tau}=H(Q^2,\mu)\int\mathrm{d}p_{L}^2
\mathrm{d}p_{R}^2\mathrm{d}kJ(p_{L}^2,\mu)J(p_{R}^2,\mu)S_{T}(k,\mu)
\delta(\tau-\frac{p_{L}^2+p_{R}^2}{Q^2}-\frac{k}{Q}),
\end{equation}
where $H(Q^2,\mu)$ is the hard function, $J(p^2,\mu)$ is the jet function
and $S_{T}(k,\mu)$ is the soft function for thrust.
$Q$ is the center-of-mass energy, $\mu$ is the renormalization
scale and $\Gamma_{0}$ is the partial width at tree-level.  
The NNLO soft function for thrust is calculated in~\cite{Schwartz:2007ib, Fleming:2007xt, Kelley:2011ng}.
The N$^3$LO jet functions are calculated in~\cite{Becher:2006qw, Becher:2010pd, Bruser:2018rad, Banerjee:2018ozf}.
The hard function can be extracted from the calculations in~\cite{Harlander:2003ai, Gehrmann:2005pd, Moch:2005tm, Gehrmann:2010ue, Gehrmann:2014vha}.
All these perturbative ingredients have been summarized in Ref.~\cite{1901.02253}.
From the factorization formula and the ingredients up to NNLO one can
derive the cumulant of the singular distributions at NNLO as
\begin{equation}
  \Gamma_{s}(x)\equiv\int_0^x dx\frac{d\Gamma_{s}}{dx}
  =\Gamma_0(1+\Gamma_s^{(1)}(x)+\Gamma_s^{(2)}(x)),
\end{equation}
where $x\equiv \tau$ or $E_X/2m_b$ are the principal variables defined above, and $\Gamma_s^{(i)}$ denotes
perturbative expansions of the partial width to the $i$-th power of the QCD coupling $\alpha_S$.
$\Gamma_s^{(i)}$ are polynomials of $\ln x$ up to order of $i$ for massive bottom quarks
and order of $2i$ for the massless cases.
We have exchanged the total radiation energies with a dimensionless variable by taking ratio to
twice of the bottom quark mass.  
Details on derivations of the singular distributions can be found in Ref.~\cite{1210.2808}.
On the other hand from conventional fixed-order calculations we can derive
the exact distribution in the three-jet region as
\begin{equation}\label{eq:3j}
  \frac{d\Gamma_{3j}(x)}{dx}
  =\Gamma_0\left(\frac{d\Gamma_{3j}^{(1)}(x)}{dx}+\frac{d\Gamma_{3j}^{(2)}(x)}{dx}\right).
\end{equation}
In a phase space slicing method, one can obtain the NNLO partial width as
\begin{equation}\label{eq:fo}
  \Gamma^{NNLO}=\Gamma_s(\delta)+\int_{\delta} dx\frac{d\Gamma_{3j}(x)}{dx},
\end{equation}
given that the cutoff parameter $\delta$ is sufficiently small and the power
corrections not accounted for can be safely neglected.
In calculating NLO corrections to the three-jet rate, we use one-loop results
of three-body decays in Ref.~\cite{1501.07226} for the Higgs boson decaying into massless
bottom quarks, and results in Ref.~\cite{hep-ph/9707448} for decaying into gluons.
In the case of decaying into massive bottom quarks, we use
GoSam~2.0~\cite{1404.7096} to generate the one-loop virtual corrections for the three-body
decay.
Reduction of loop integrals is performed with Ninja~\cite{1203.0291,1403.1229}
and scalar integrals are calculated with OneLOop~\cite{0903.4665,1007.4716}.
Further we can define the following damping factor,
\begin{align}
\mathcal{D}(x)=&\exp{[\Gamma^{(1)}_s(x)+\Gamma^{(2)}_s(x)-(\Gamma_s^{(1)}(x))^2/2}]
\nonumber\\
\equiv &\exp{[D^{(1)}(x)+D^{(2)}(x)]},
\end{align}
which vanishes as $x$ goes to zero and can be expanded perturbatively in the three-jet
region.
Based on the damping factor we can construct the following damped predictions by
reweighting the fixed-order distributions in Eq.~(\ref{eq:3j}),
\begin{equation}\label{eq:damp}
  \Gamma^{damp.}=\int dx\mathcal{D}(x)\left(\frac{d\Gamma_{3j}(x)}{dx}-
  D^{(1)}(x)\Gamma_0\frac{d\Gamma^{(1)}_{3j}}{dx}\right).
\end{equation}
It is not difficult to show that above integration can reproduce the partial
width up to NNLO.
We point out that even though we write Eq.~(\ref{eq:damp}) as an explicit integration
over $x$ the reweighting can be applied at the exclusive level for the full phase space
of three-jet production, where $x$ can be reconstructed on the event-by-event basis.
We use a slightly modified version of the damped three-jet distributions as
the matrix element inputs for POWHEG matching.
As shown in Appendix~\ref{sec:appa} such a distribution agrees with the conventional NLO
result in the resolved three-jet region while being normalized to the exact NNLO partial width
upon integration over the full three-jet phase space.
The damping factors used are inspred by the Sudakov factors in the MiNLO method~\cite{1912.09982}
though they are essentially different.
In $\mathcal{D}(x)$ we always fix the QCD scale to the hard scale of $\sim m_H$ rather
than using a dynamic scale varying with $x$.
Also for the massive case in the exponent there only exist single logarithms of $x$
accompanied by logarithms of the mass of the bottom quark.
As will be explained in the following section, introduction of the damping
factor is to ensure the decay rate integrable over the full born phase space, and
realize an unitarized seperation of the two and three-jet samples in the POWHEG matching.
The Sudakov resummation in our approach is completely from the
shower MCs, unlike in the MiNLO method.

\subsection{Parton shower matching}

The POWHEG method has been widely used for matching NLO computations with
parton showers.
The basic idea is to single out all different singular regions in the real
emissions and to parametrize the phase space with the so-called underlying
Born phase space and additional radiation variables.
A probability distribution fully differential in the Born phase space
is calculated. 
Meanwhile a Sudakov factor that describes the no radiation
probability as a function of radiation $k_T$ at each configuration of the Born
phase space is also constructed.
The Sudakov factor is calculated based on the exact matrix elements at NLO.
With the two successive probability distributions, a first hard emission is generated
by POWHEG from the Born phase space, and later emissions as handled by parton shower can
only be allowed for $k_T$ lower than that of the first emission.
Here we reproduce the POWHEG formula for the case where only final state emissions
are presented~\cite{0709.2092},
\begin{equation}
  d\Gamma=\bar B({\Phi}_n)d{\Phi}_n\Big(\Delta({\Phi}_n,p_T^{min})
  +\Delta({\Phi}_n,k_T({\Phi}_{n+1}))\frac{R({\Phi}_{n+1})}
  {B({\Phi}_n)}\mathrm{d}{\Phi}_{rad}\Big),
\end{equation}
where $\Phi_n$ and $\Phi_{rad}$ are the Born phase space and the radiation phase space
respectively, $p_T^{min}$ is a small cutoff for the transverse momentum of the POWHEG hard emission.
The $\bar B({\Phi}_n)$ can be thought as projection of the NLO distributions
on to the Born phase space and is given by
\begin{align}
\bar B({\Phi}_n)&=[B({\Phi}_n)+V({\Phi}_n)]\nonumber\\
  &+\int\mathrm{d}{\Phi}_{rad}[R({\Phi}_{n+1})-C({\Phi}_{n+1})],
\end{align}
for a typical subtraction method at NLO like FKS method~\cite{hep-ph/9512328}, with $R$ and $C$ being
the QCD corrections from real emissions and the counterterms respectively.
Precise definition of the Sudakov form factor is given by
\begin{equation}
  \Delta({\Phi}_n,p_T)=\exp\Big(-\int\frac{[\mathrm{d}{\Phi}_{rad}R({\Phi}_{n+1})
  \theta(k_T({\Phi}_{n+1})-p_T)]^{\bar{{\Phi}}_n={\Phi}_n}}
  {B({\Phi}_n)}\Big).
\end{equation}
Further details on the POWHEG framework can be found in Ref.~\cite{0709.2092}.
In standard MC programs, like PYTHIA8~\cite{Sjostrand:2014zea}, there are similar approaches implementing
the so-called matrix element reweighting~\cite{hep-ph/0611247}.
The difference is that $\bar B(\Phi_n)$ is replaced by the tree-level matrix element
$B(\Phi_n)$.
Our matched calculations start with the NLO calculations for three-body decays
of the Higgs boson, namely with tree-level processes being $H\rightarrow b\bar b g$,
$q\bar q g$ for decays into
bottom quarks or light quarks, and $H\rightarrow ggg(q\bar q)$ for decays into
gluons.
One can not simply integrate over the full Born phase space $\Phi_3$ since the
tree-level matrix elements have further soft or collinear singularities.
We apply the reweighting procedure designed in Eq.~(\ref{eq:damp}) to suppress
those singularities and render the $\bar B(\Phi_3)$ kernel
integrable in the full phase space, denoted as $\bar B^*(\Phi_3)$, with
the precise definition given by
\begin{align}
  \bar B^*({\Phi}_3)&=\mathcal{\tilde{D}}(x(\Phi_3))
  [B({\Phi}_3)(1-\tilde{D}^{(1)}(x(\Phi_3)))+V({\Phi}_3)]\nonumber\\
  &+\int\mathrm{d}{\Phi}_{rad}[\mathcal{\tilde{D}}(x(\Phi_{4}))R({\Phi}_{4})
  -\mathcal{\tilde{D}}(x(\Phi_3))C({\Phi}_{4})].
\end{align}
Note the principal variable used in the damping factor $\mathcal{\tilde{D}}(x)$ is
constructed on the basis of infrared and collinear safe observables, $E_X$
for the massive case and $\tau$ for the massless cases. 
We further split the Born phase space and the contributions to $\bar B^*$ into
the resolved three-jet region and the unresolved one using the $\tau$ variable
constructed from $\Phi_3$ for both the massive and massless cases, and a merging parameter $\tau_{m}$, 
\begin{align}
  d\Gamma &=\bar B^*({\Phi}_3)\theta(\tau(\Phi_3)-\tau_{m})
  d{\Phi}_3\Big(\Delta({\Phi}_3,p_T^{min})
  +\Delta({\Phi}_3,k_T({\Phi}_{4}))\frac{R({\Phi}_{4})}
  {B({\Phi}_3)}\mathrm{d}{\Phi}_{rad}\Big)\nonumber\\
  &+\bar B^*({\Phi}_3)\theta(\tau_{m}-\tau(\Phi_3))d{\Phi}_3.
\end{align}
We choose $\tau_m$ to be larger than the position where the Sudakov peak locates.
Based on above distinctions we feed the POWHEG events with $\tau(\Phi_3)>\tau_{m}$
to PYTHIA8 for parton shower as in the nominal matching calculations at NLO.
That consists of our three-jet samples.
For events with $\tau(\Phi_3)<\tau_{m}$ we replace them with randomly generated
events from PYTHIA8 for the same hadronic decay channel of Higgs boson, based on
parton shower with native matrix elements reweigthing.\footnote{The first-order
matrix elements reweighting was implemented in PYTHIA8 for the Higgs boson decaying into quarks
but not for decaying into gluons.} 
In this category after parton shower we calculate $\tau$ values
for splittings from the two daughters of the Higgs boson based on the
MC record.
We require the larger one of the two $\tau$ values is smaller
than $\tau_m$.
Otherwise we repeat the shower.
That generates our two-jet samples.
With the shower veto we can minimize the overlapping of phase space probed 
in the three-jet samples and the two-jet samples.
In our merging scheme the phase-space overlapping can not be avoided since neither
the parton shower nor POWHEG emissions are ordered in $\tau$.
In the end we will vary the merging scale $\tau_m$ and count the variations
as part of the perturbative uncertainties.
The total normalization of our matched results is fixed to the exact
NNLO partial width by construction.
Our results have the leading logarithmic accuracy
same as PYTHIA8 in the Sudakov region,
and maintain the NLO matched
with parton shower accuracy in region with three hard jets.
We note in the case of massless quarks, the MiNLO method in Ref.~\cite{1912.09982}
provides a smooth matching of the two-jet and three-jet samples without
introducing an explicit merging scale.
Distributions in the Sudakov region have been reweighted according to
results from analytical resummations.
Our calculations for the Higgs boson decaying into massless quarks
and gluons can be compared with those in Ref.~\cite{2009.13533}.
First in the GENEVA method the decays are seperated into 2, 3
and 4-jet contributions with cuts on the 2 and 3-jettiness,
${\mathcal T}^{{\rm cut}}_2$ and ${\mathcal T}^{{\rm cut}}_3$ respectively.
Note in Higgs decays the 2-jettiness is equivalent to the thrust
variable used here, $\tau={\mathcal T}_2/(2m_H)$. 
In the 2-jet contributions, the total rate is calculated at NNLO,
and is further improved with analytic resummations on ${\mathcal T}_2$.
The 3-jet contributions are based on predictions of analytic resummations
on ${\mathcal T}_2$ matched with NLO fixed-order calculations.
The 3-jet and 4-jet contributions also include resummations on
${\mathcal T}_3$.
Once matched with parton showers, by choosing a small ${\mathcal T}^{{\rm cut}}_2\,(\sim 1\,{\rm GeV})$ and a dedicated treatment on the first emission, the precisions of
analytical resummations on ${\mathcal T}_2$ are maintained.
In our approach we choose a much larger value of $\tau_m$ that is beyond the
Sudakov peak.
Thus the two-jet samples are dominant.
Below $\tau_m$ the resummations are realized by the Sudakov factors in
parton showers.

\section{Numerical results}\label{sec:num}
In the numerical calculations we set the mass of the Higgs boson $m_H=125.09\,{\rm GeV}$,
the vacuum expectation value $v=246.22\,{\rm GeV}$, and $\alpha_S(M_Z)=0.1181$~\cite{Tanabashi:2018oca}.
The pole masses of the top quark and bottom quark are set to $172.5\,{\rm GeV}$
and $4.78\,{\rm GeV}$ respectively.
We use the 3-loop running of $\alpha_S$ and of the bottom quark mass in a
$\overline {\rm MS}$ scheme with 5 light flavors~\cite{hep-ph/0004189}.
We set the renormalization scale to the mass of the Higgs boson unless specified.
We use POWHEG-Box-v2~\cite{1002.2581} for matching of matrix elements and PYTHIA8.2~\cite{Sjostrand:2014zea}
for final state QCD showers and hadronizations with the Monash
tune~\cite{Skands:2014pea}.
We turn off multiple particle interactions and decays of $B$ hadrons
for simplicity.
The generated MC event samples are analysed with Rivet program~\cite{1003.0694}.

\subsection{Partial width at fixed order}
\begin{figure}[ht]
\centering
\includegraphics[width=0.48\textwidth]{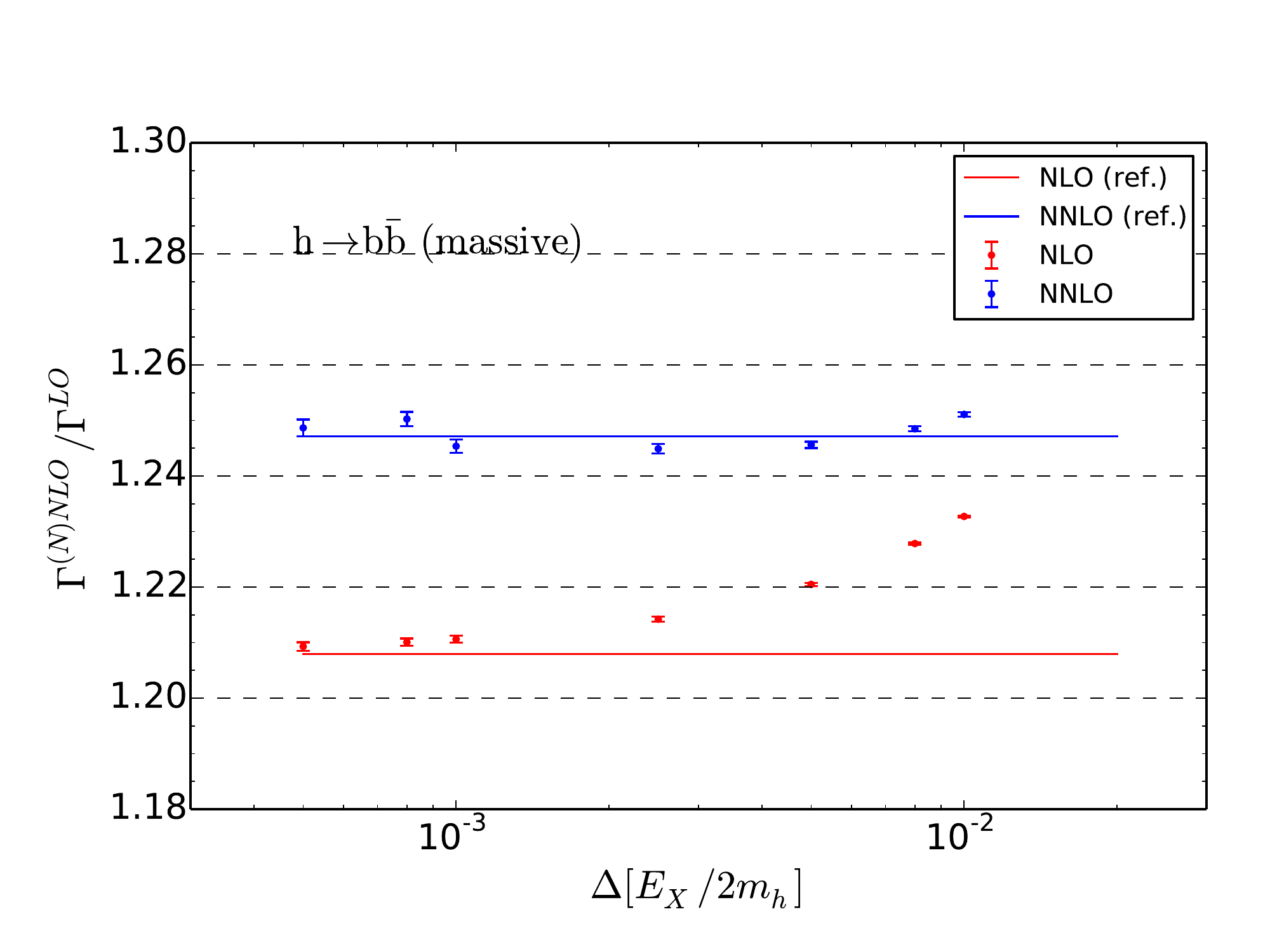}
\caption{
Dependence of the total partial width calculated at NLO and NNLO on the
phase space slicing parameter for the Higgs boson decaying into bottom
quarks with full bottom-quark mass dependence.
The phase space slicing variable is chosen to be the total radiation energies
normalized to twice of the Higgs boson mass.
The horizontal lines represent corresponding reference predictions from
the literature, see text for details.
All predictions are normalized to the LO partial width.
\label{fig:fototbb}}
\end{figure}

We start with results on the total partial decay width of the Higgs
boson decaying into bottom quarks and gluons up to NNLO in QCD.
For bottom quarks we show separately results with full bottom quark mass
dependence and results neglecting bottom quark mass except in the
Yukawa coupling.
In Fig.~\ref{fig:fototbb} we plot dependence of the NLO and NNLO width
on the cutoff of the phase space slicing variable, which is chosen to be the total
radiation energies normalized to twice of the Higgs boson mass,
for the case of massive bottom quarks.
All predictions have been normalized to the LO partial width for simplicity while
the absolute values will be summarized in later section.
The scattering points with error bar represent our calculations with MC
statistical uncertainties.
They clearly approach the genuine NLO/NNLO predictions taken from Ref.~\cite{1911.11524}
that are represented by the horizontal lines.
In general the differences due to the neglected power corrections are
within the MC errors which are about one per mille, if the cutoff is
below $10^{-3}$.

\begin{figure}[ht]
\centering
\includegraphics[width=0.48\textwidth]{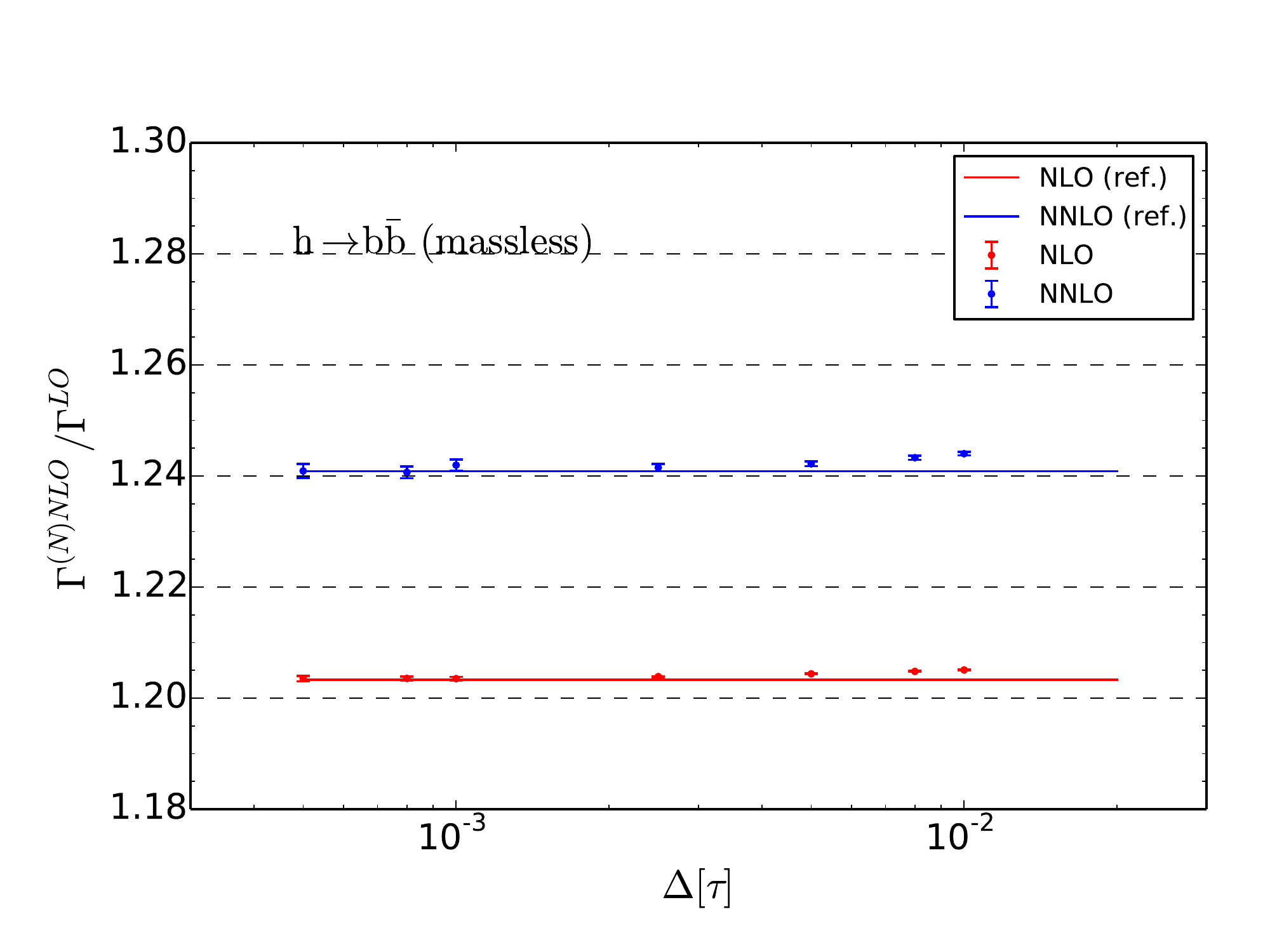}\hspace{0.1in}
\includegraphics[width=0.48\textwidth]{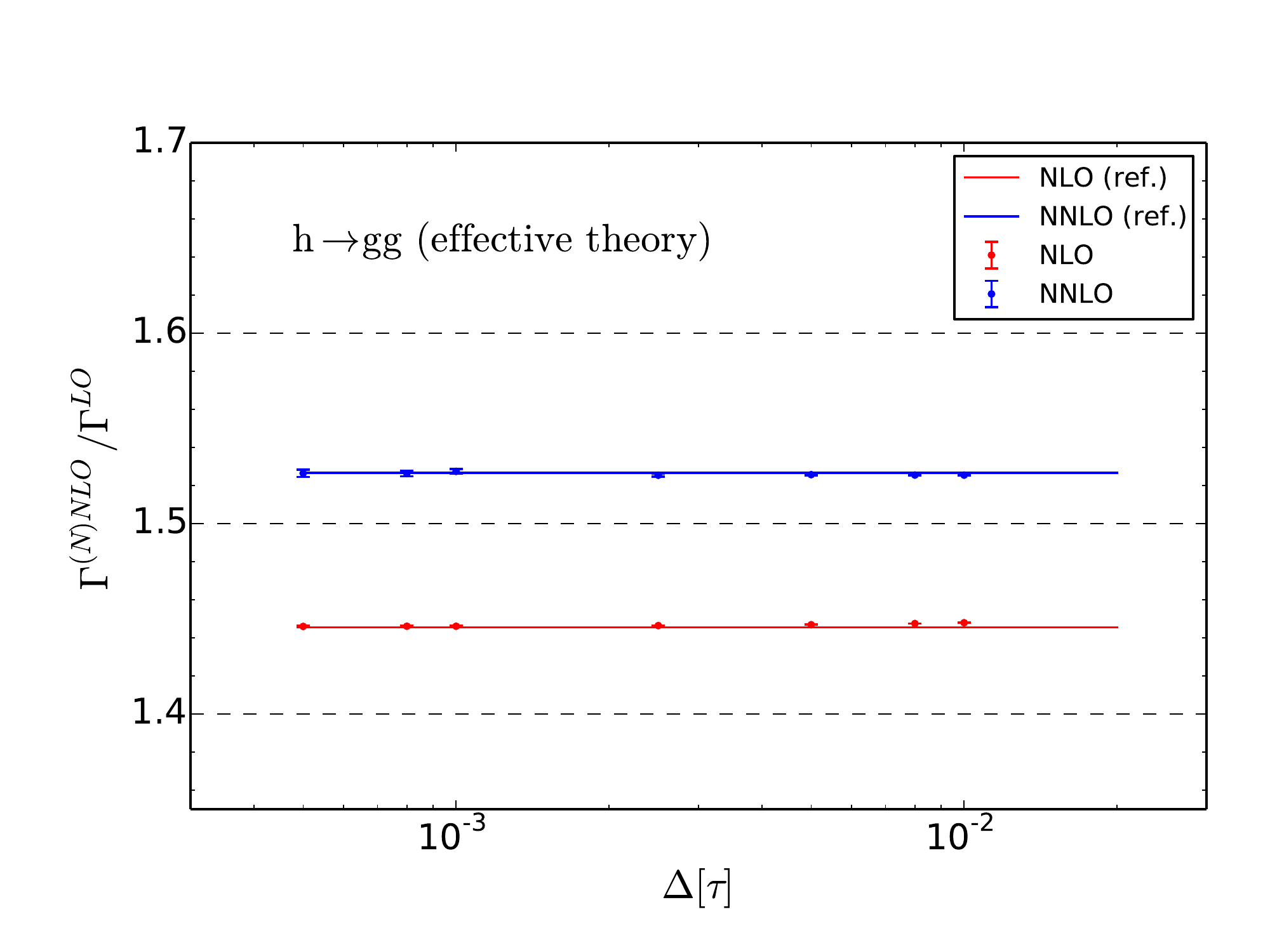}
\caption{
Similar to Fig.~\ref{fig:fototbb} for the Higgs boson decaying into massless bottom
quarks and gluons.
The phase space slicing variable is chosen to be $\tau$ variable constructed
from all final state partons.
\label{fig:fototgg}}
\end{figure}

In Fig.~\ref{fig:fototgg} we present similar results for the Higgs boson decaying
into massless bottom quarks and gluons.
In this case the phase space slicing variable is the $\tau$ variable constructed
from all final state partons.
The reference values of NLO/NNLO width are taken from Ref.~\cite{1707.01044}
for bottom quarks and gluons.
Again we see excellent agreements between our numerical results and the analytical
results in the literature.
We find a cutoff parameter of the order $\sim 10^{-2}$ is good enough to
ensure the power corrections being negligible especially for the gluon channel.
We can further compare the $K$-factors of the bottom-quark channel in the calculations
with full mass dependence and using massless approximation.
At both NNLO and NLO the $K$-factors with respect to LO are larger by about half
percent in the calculation with full mass dependence.
\begin{table}[h!]
\centering
\begin{tabular}{|c|ccccc|}
\hline
[MeV/GeV] & $\Gamma_{b\bar b}(m_b\ne 0)$ & $\Gamma_{b\bar b} (m_b=0)$ & $\,\,\Gamma_{gg}\,\,$ 
  & $\,\, \alpha_S(\mu)\,\,$ & $\,\,m_b(\mu)\,\,$ \tabularnewline
\hline                      
  $\mu=m_h/2$  &2.314& 2.320  &0.3488  & 0.1252 & 2.914 \tabularnewline
\hline                      
  $\mu=m_h$  &2.293& 2.302  & 0.3437 & 0.1127 & 2.744 \tabularnewline
\hline                      
  $\mu=2m_h$  &2.252& 2.263 & 0.3290 & 0.1025 & 2.601 \tabularnewline
\hline
\end{tabular}
\caption{
 Partial width for the Higgs boson decaying into bottom quarks and gluons calculated
 at NNLO for different choices of the renormalization scales.
 The last two columns show the running QCD coupling and bottom quark mass used
 in the calculations.
\label{tab:inc}}
\end{table}

We summarize the partial decay width at NNLO for the bottom-quark channel and
gluon channel in Table.~\ref{tab:inc}.
The results are calculated with our numerical program and further cross checked
with available calculations in the literatures~\cite{1707.01044,1911.11524}.
We also include values of the running QCD coupling and the bottom quark mass at the
chosen renormalization scales.
With the full bottom-quark mass dependence the NNLO width is reduced by about
half percent due to suppressions of the phase space.
Here and below we have not included interference contributions between
the two operators in Eq.~(\ref{eq:leff}) which can be calculated separately.
\begin{figure}[ht]
\centering
\includegraphics[width=0.48\textwidth]{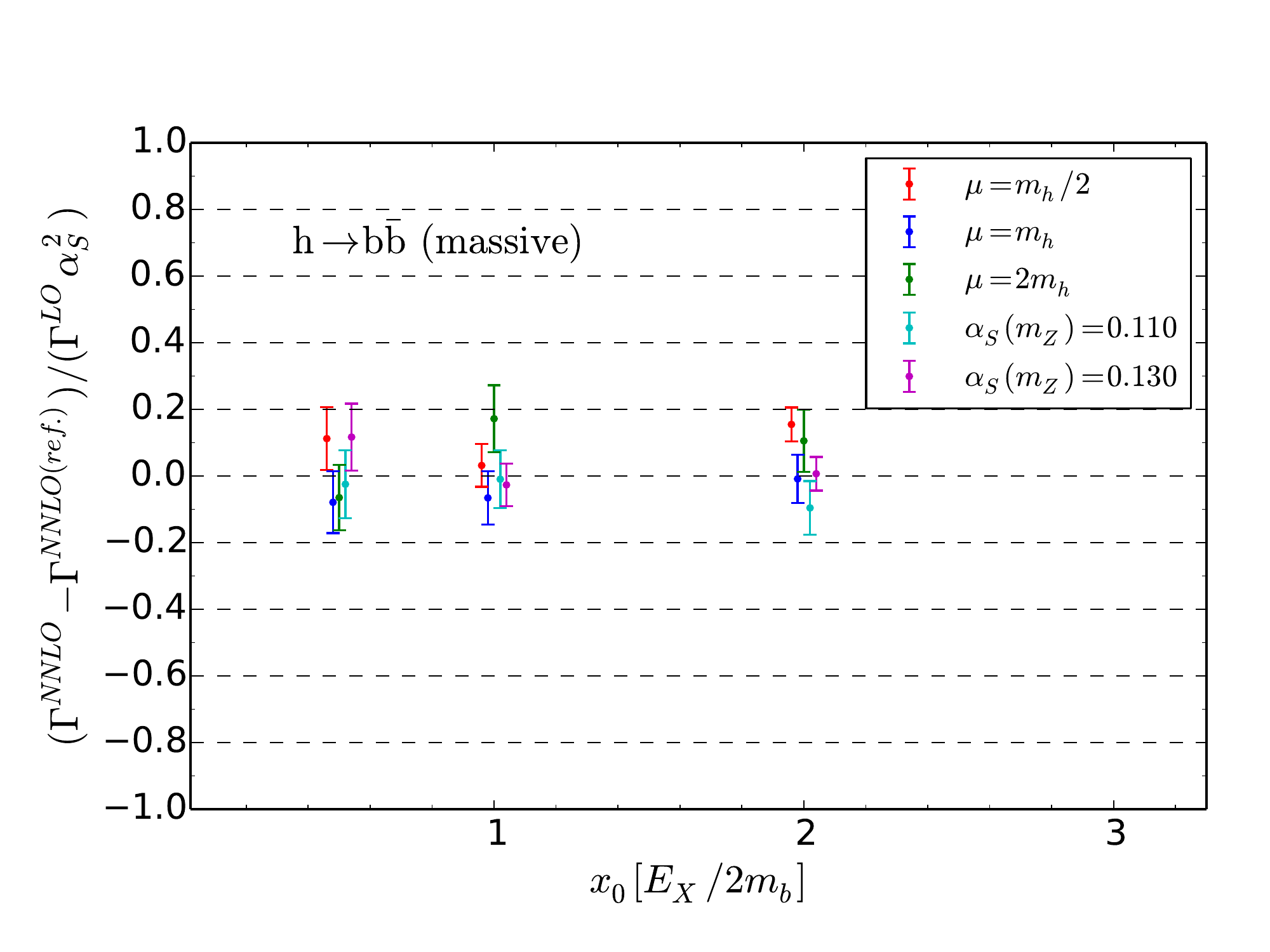}\hspace{0.1in}
\includegraphics[width=0.48\textwidth]{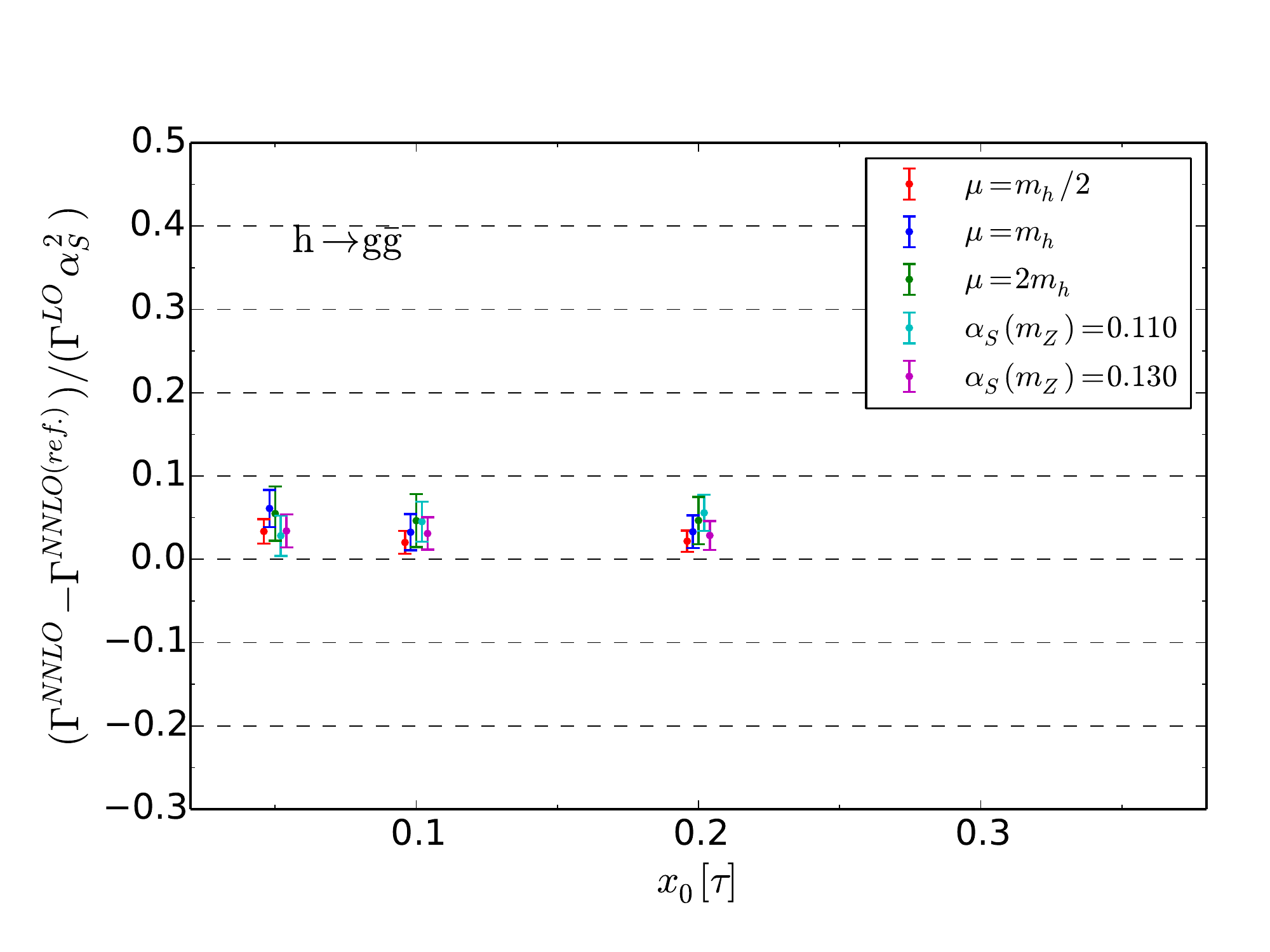}
\caption{
Validations on reproduction of the partial width at NNLO with the damping
procedure outlined, for the Higgs boson decaying into bottom quarks or gluons,
with different choices of the damping parameter.
Results with MC errors are compared to the reference predictions from
literatures for difference choices of the QCD coupling and the
renormalization scale.
\label{fig:damtotbb}}
\end{figure}

\begin{figure}[ht]
\centering
\includegraphics[width=0.48\textwidth]{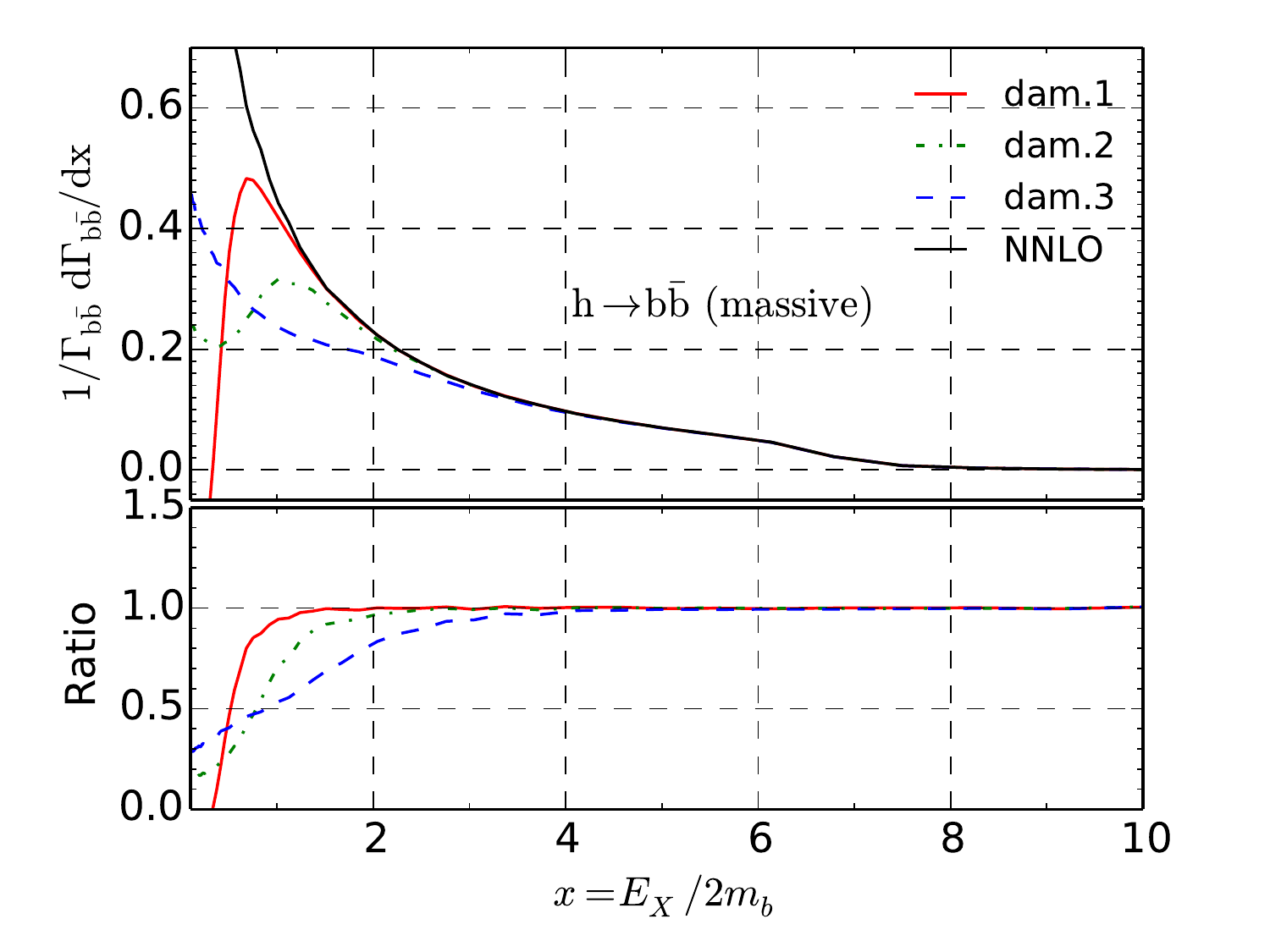}\hspace{0.1in}
\includegraphics[width=0.48\textwidth]{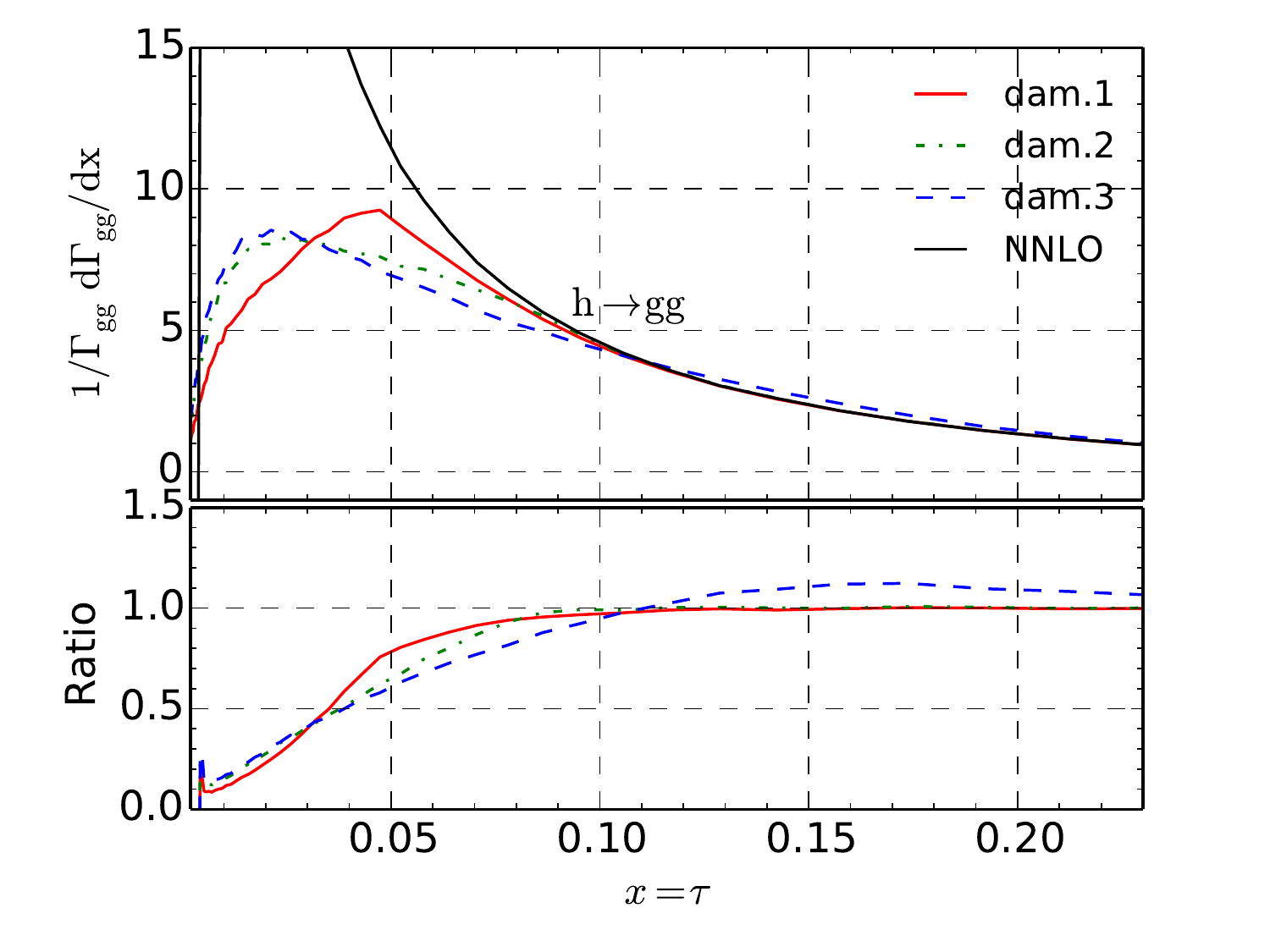}
\caption{
Distributions of the total radiation energies ($\tau$ variable),
comparing the fixed-order predictions
and various damped results, for the Higgs boson decaying into bottom quarks (gluons).
\label{fig:dampdisbb}}
\end{figure}

We move to the damped predictions from integrating over the three-body phase space.
We demonstrate the effectiveness of the procedure by comparing the damped
predictions with the exact NNLO predictions for different choices of the
renormalization scale, $\mu=\{m_h/2,\,m_h,\,2m_h\}$, and the QCD coupling constant,
$\alpha_S(m_Z)=\{0.110,\,0.1181,\,0.130\}$.
Fig.~\ref{fig:damtotbb} shows the comparison for the total partial
width of the Higgs boson decaying into bottom quarks with full mass dependence,
and of the Higgs boson decaying into gluons.
We include results for three different choices of the
damping parameter, $x_{0}=\{0.5,\,1,\,2\}$ for the case of massive bottom quarks
with $x=E_X/2m_b$,
and $x_{0}=\{0.05,\,0.1,\,0.2\}$ for the case of gluons with $x=\tau$.
Precise definition of the damping parameter is given in Appendix~\ref{sec:appa}.
Differences between the damped NNLO predictions and the exact NNLO predictions
have been normalized to the leading order width times $\alpha_S^2$.
The damped calculations reproduce the exact results to a precision of one
per mille in general.
In Fig.~\ref{fig:dampdisbb} we further compare the exact and damped predictions
on the distribution of radiation energies $E_X$ ($\tau$ variable) for decaying
into bottom quarks (gluons).
We note that for a NNLO calculation of the total partial width it only
predicts above distributions at NLO.
The distributions from exact calculations diverge when $E_X$ or $\tau$ goes
to zero due to the soft and collinear singularities.
The damped predictions render a strong suppression in the soft and collinear
region and ensure the distributions being smoothly integrable
in the full phase space.

In the following calculations when matching the NNLO calculations with parton
shower we use a damping parameter of $x_{0}=0.5(0.05)$ for the massive (massless)
case.
That ensures the total partial width normalized automatically to the NNLO
fixed-order predictions and also in resolved three-jet region the underlying parton-level
results agree with the NLO fixed-order predictions.

\subsection{Predictions at parton level}
\begin{figure}[ht]
\centering
\includegraphics[width=0.32\textwidth]{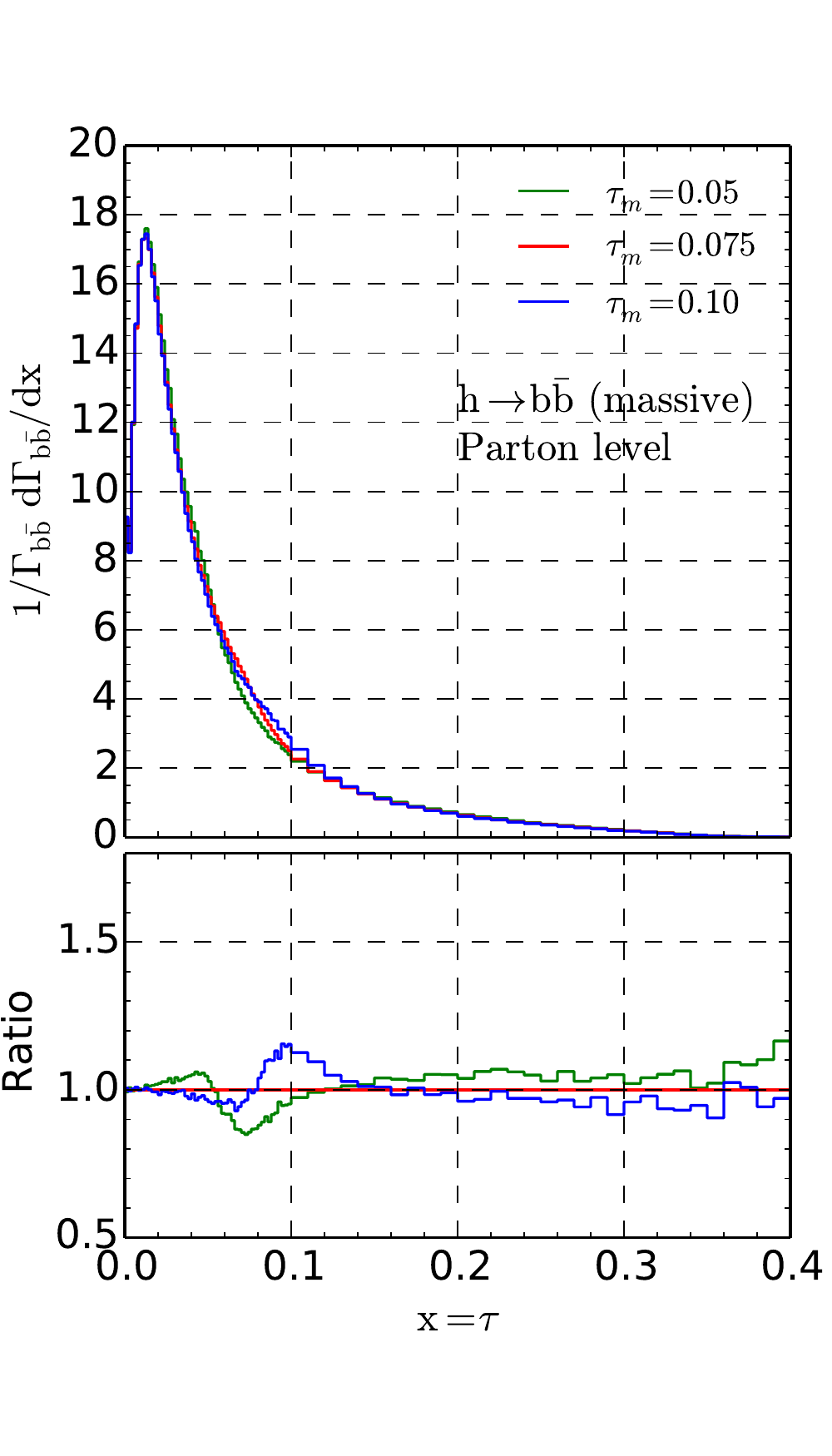}
\includegraphics[width=0.32\textwidth]{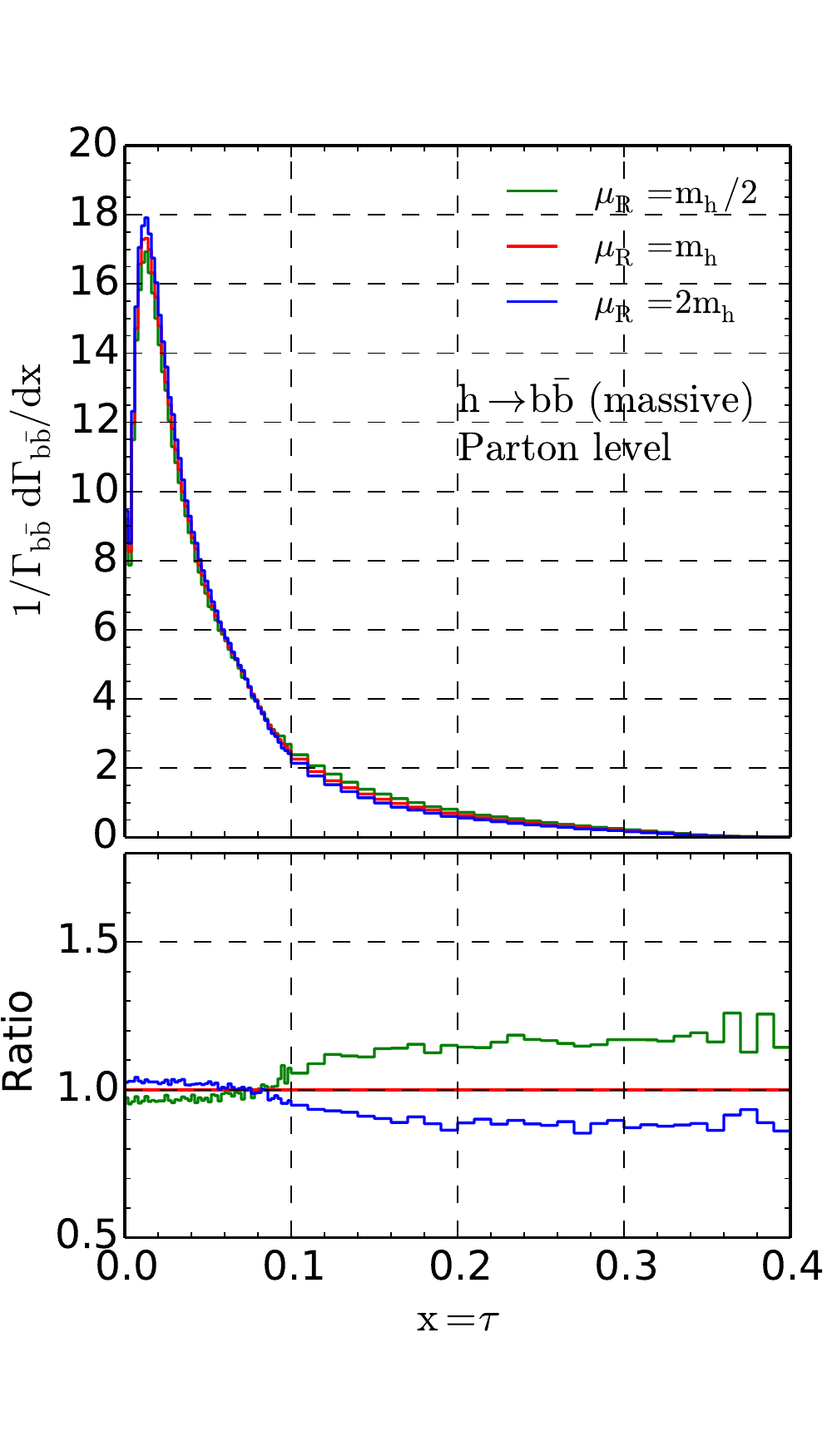}
\includegraphics[width=0.32\textwidth]{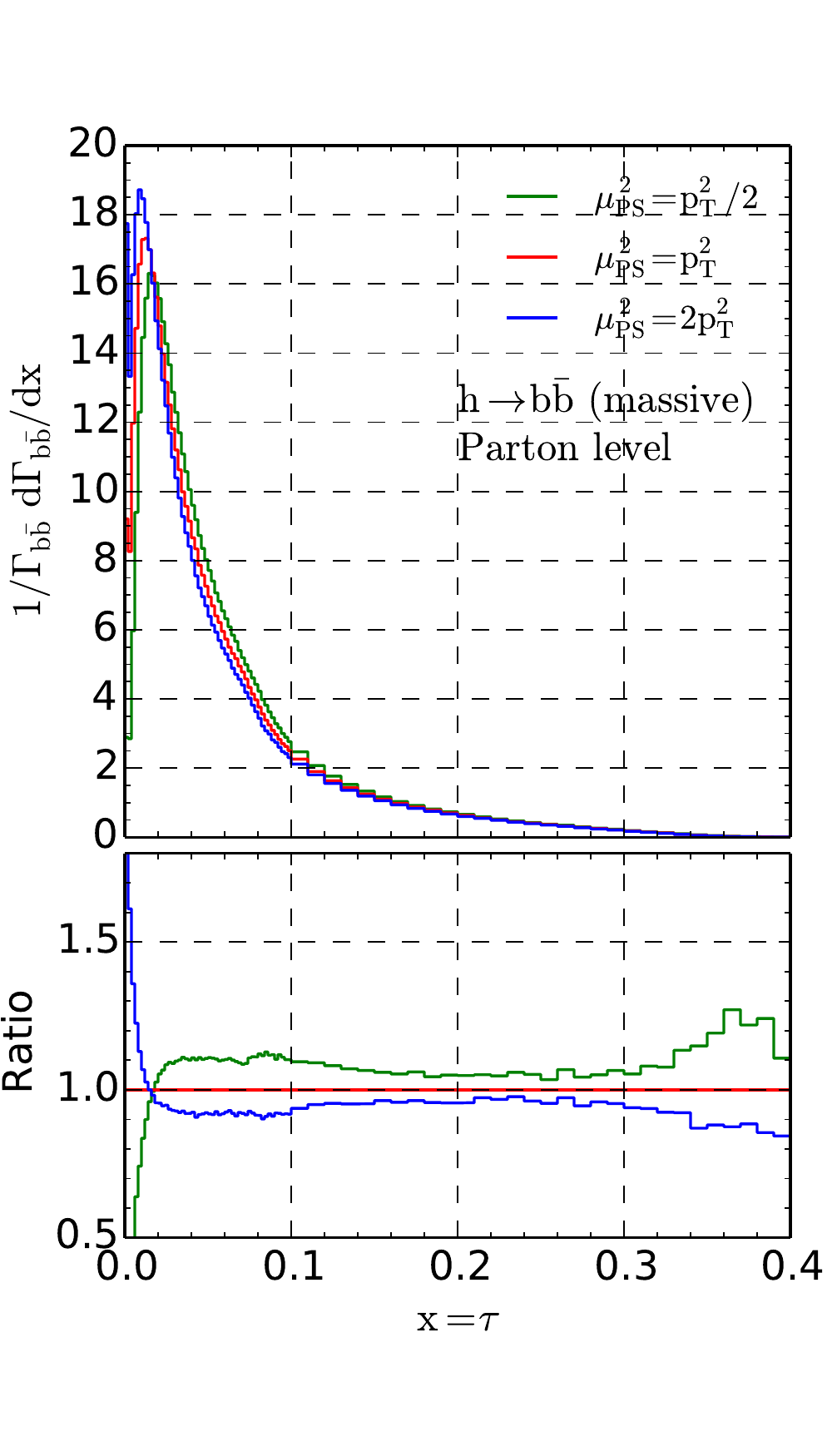}
\caption{
$\tau$ distribution calculated at parton level matched with parton shower
for the Higgs boson decaying into bottom quarks.
The three plots show dependence of the predictions on the merging scale,
the renormalization scale and the scale in parton shower.
In each plot the upper panel shows the normalized distribution
and the lower panel shows ratios of different predictions.
\label{fig:sysbb}}
\end{figure}

We proceed with predictions of matching the NNLO fixed-order calculations with
parton shower using the POWHEG method and the merging scheme outlined in earlier
sections.
We have checked the predictions for a variety of event shape variables but only
show here results for $\tau$ variable due to limited space.
In Fig.~\ref{fig:sysbb} we show our nominal predictions for the Higgs boson decaying
into massive bottom quarks with the default choice of
the merging scale $\tau_{m}=0.075$, the renormalization scale $\mu_R=m_h$,
and the scale in parton shower $\mu_{PS}=p_T$.
From left to right, we show separately the variations due to changes
of above three scales in turn.
In each plot the upper panel shows the normalized distribution
and the lower panel shows ratios of different predictions.
We notice that even there are massive particles in the final state, the thrust or
$\tau$ variable can be constructed in a similar way as in the massless case.

The normalized $\tau$ distribution shows a clear dependence on the
merging scale in the designed merging region around $\tau\sim 0.075$.
For alternative merging scales of $\tau_{m}=0.05, 0.10$, the
normalized distribution can change by 15\% at most because of the
differences of the three-jet and two-jet samples in the merging region.
However, when going away from the merging region, the distributions
are less affected.
The change of renormalization scale shows its impact in the region of $\tau$
larger than the merging scale.
Variation of the renormalization scale by a factor of two leads to
a change of the distribution by about 15\% in that region.
In the two-jet region the normalized distributions vary by only a few
percents in order to maintain the total partial width to the NNLO
values at the associated scales.
The variation of shower scale can shift the height and position of the peak
of the distribution significantly that are determined by the two-jet sample.
Changing the square of the shower scale by a factor of two serve as an estimate of the
uncertainties due to QCD resummation in the region of $\tau$ below the merging
scale.
We also notice a non-negligible change of the distribution for $\tau$ well
above the merging scale.
That is because in the three-jet sample after the first hard emission from
POWHEG the subsequent radiations are handled by parton shower and are
thus affected by the choice of shower scale.
The impact of these subsequent emissions is especially pronounced in the
tail region of $\tau$ where fixed-order predictions are limited due to
phase space constraints.
\begin{figure}[ht]
\centering
\includegraphics[width=0.32\textwidth]{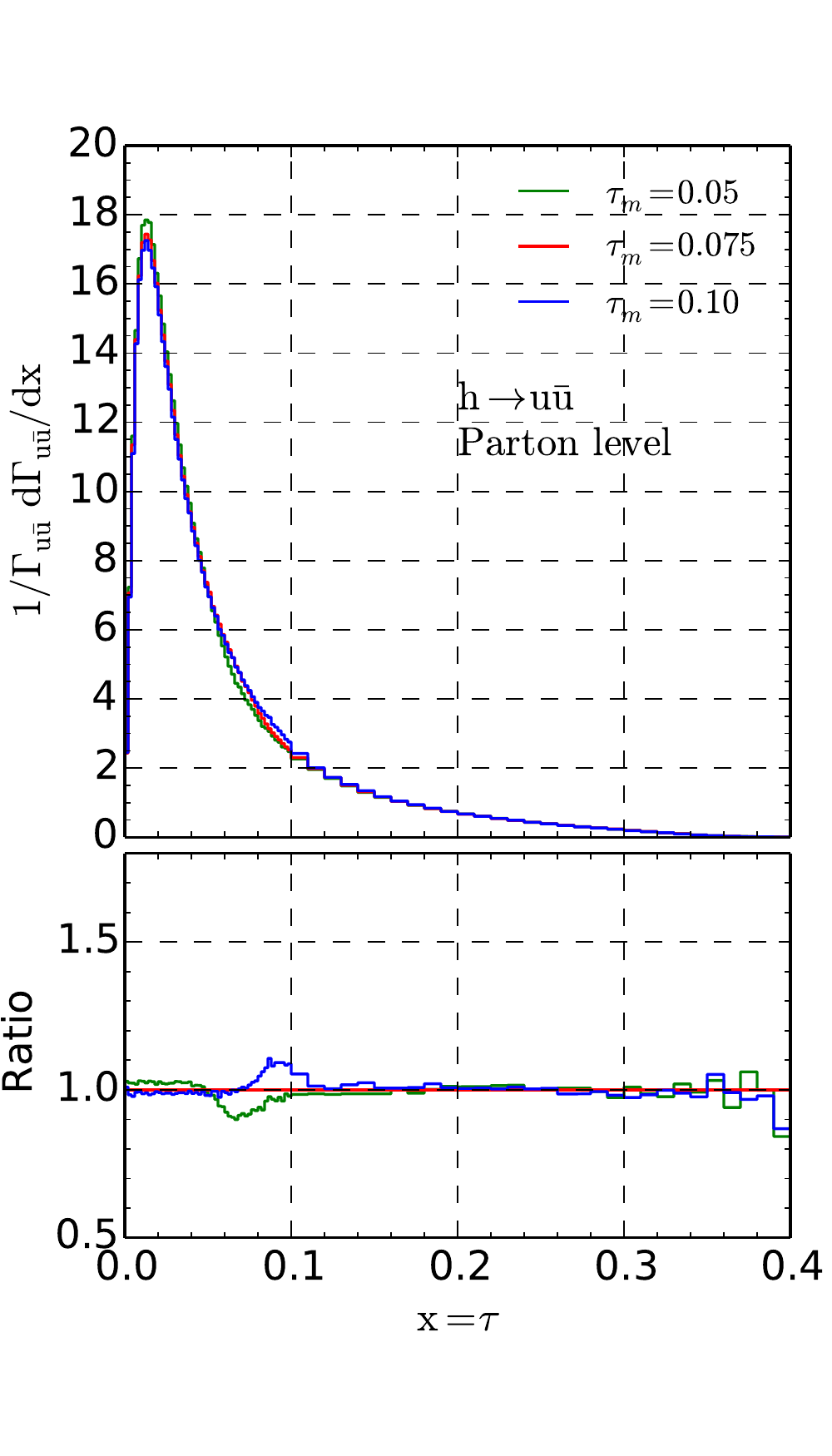}
\includegraphics[width=0.32\textwidth]{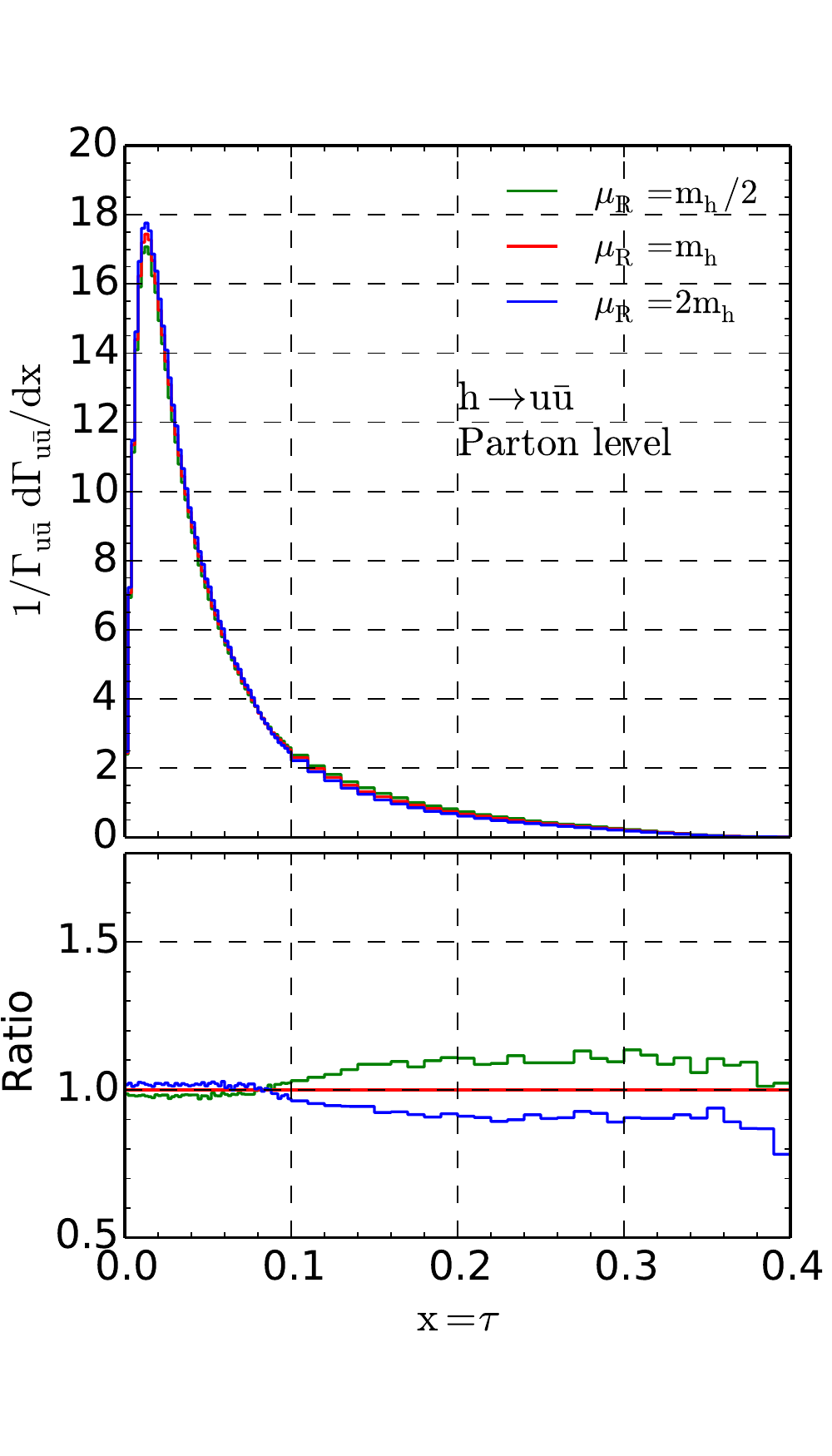}
\includegraphics[width=0.32\textwidth]{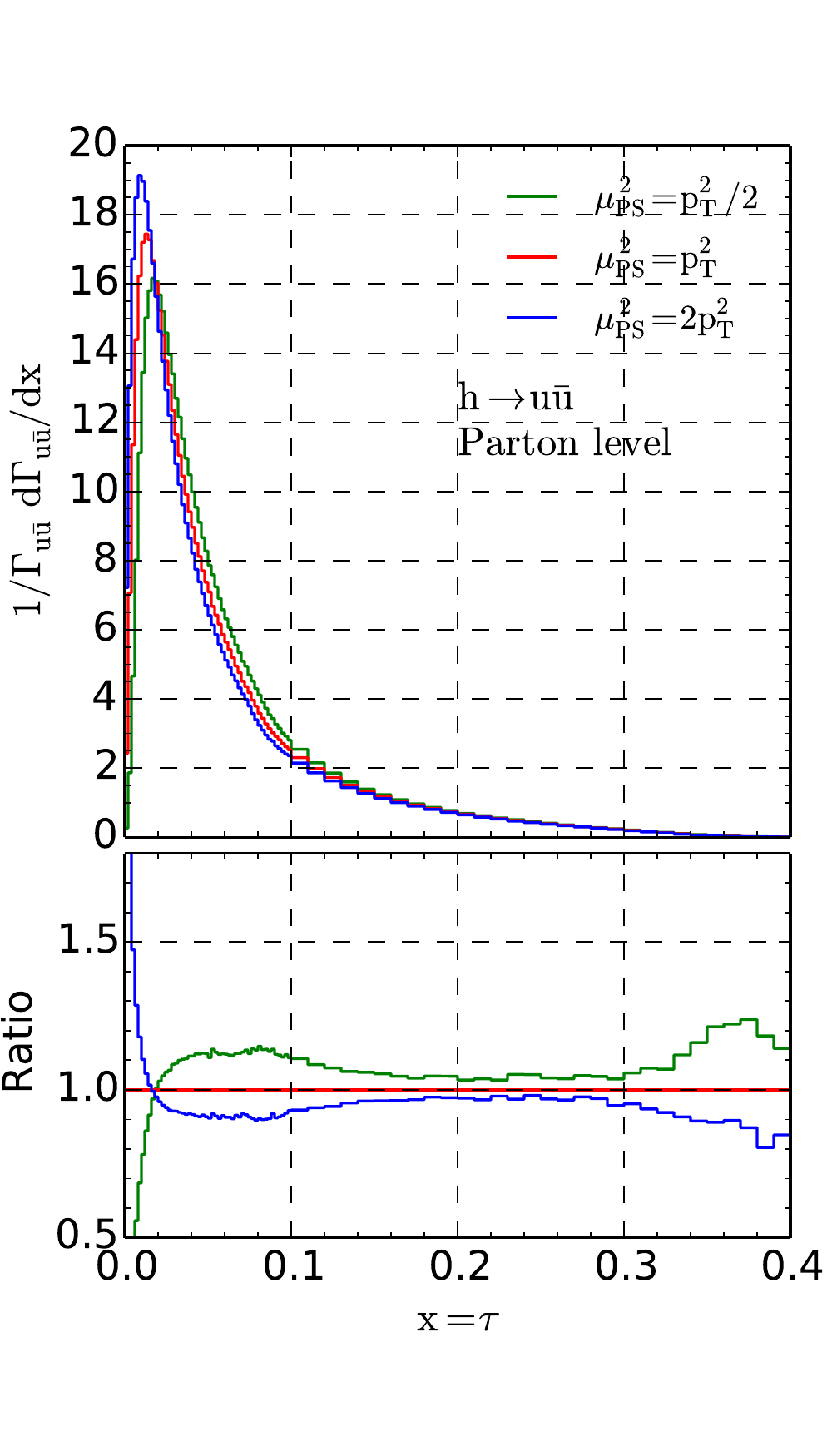}
\caption{
Similar to Fig.~\ref{fig:sysbb} for $\tau$ distributions for the Higgs boson decaying
into up quarks. 
\label{fig:sysuu}}
\end{figure}

For our NNLO calculation of the Higgs boson decaying into massless bottom
quarks, in principle we can also match it with parton shower if neglecting
the mismatch of the bottom quark mass used in matrix element calculations
and that used in parton shower.
Here we instead apply our massless calculations to the Higgs boson decaying into light
quarks, for example, strange quarks or even up and down quarks.
In such cases both quark masses in the matrix element calculations and in the parton showers
can be safely neglected.
We show results for the Higgs boson decaying into up quarks in below since the matched
predictions will be independent of the flavor of quarks at the parton level.
In Fig.~\ref{fig:sysuu} we plot the $\tau$ distribution and its dependence on the
three scales in a same format as Fig.~\ref{fig:sysbb}.
The genuine feature is similar to that of the Higgs boson decaying into bottom quarks and
we only highlight a few differences comparing to the case of bottom quarks.
The normalized distribution shows less dependence on the merging scale in full range
of $\tau$.
For instance, the variations are at most 10\% in the merging region and are
negligible for $\tau$ above the merging scales.
The distribution also show a slightly smaller variation on the renormalization
scale possibly due to mass dependence in the POWHEG matching method.
\begin{figure}[ht]
\centering
\includegraphics[width=0.32\textwidth]{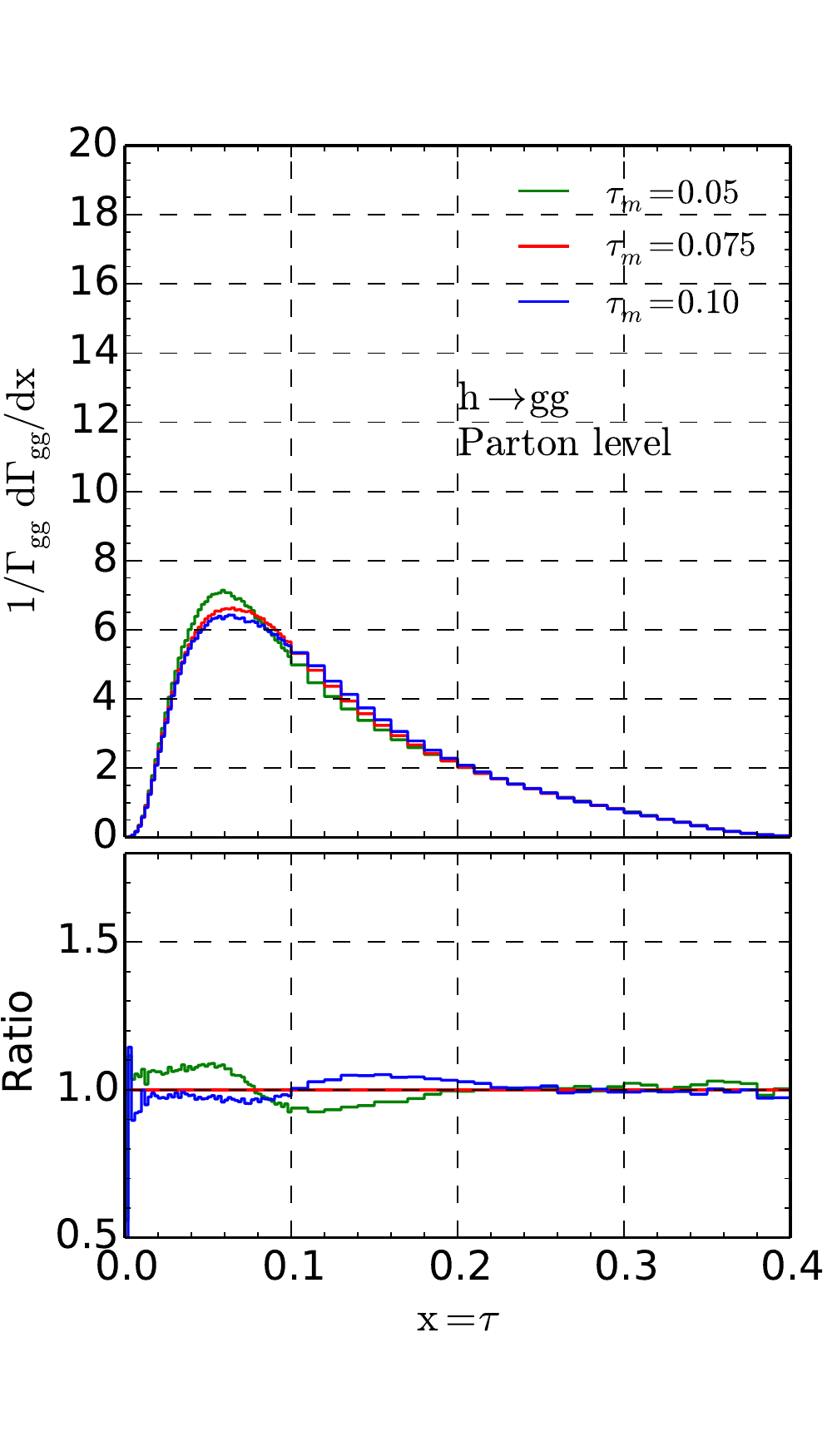}
\includegraphics[width=0.32\textwidth]{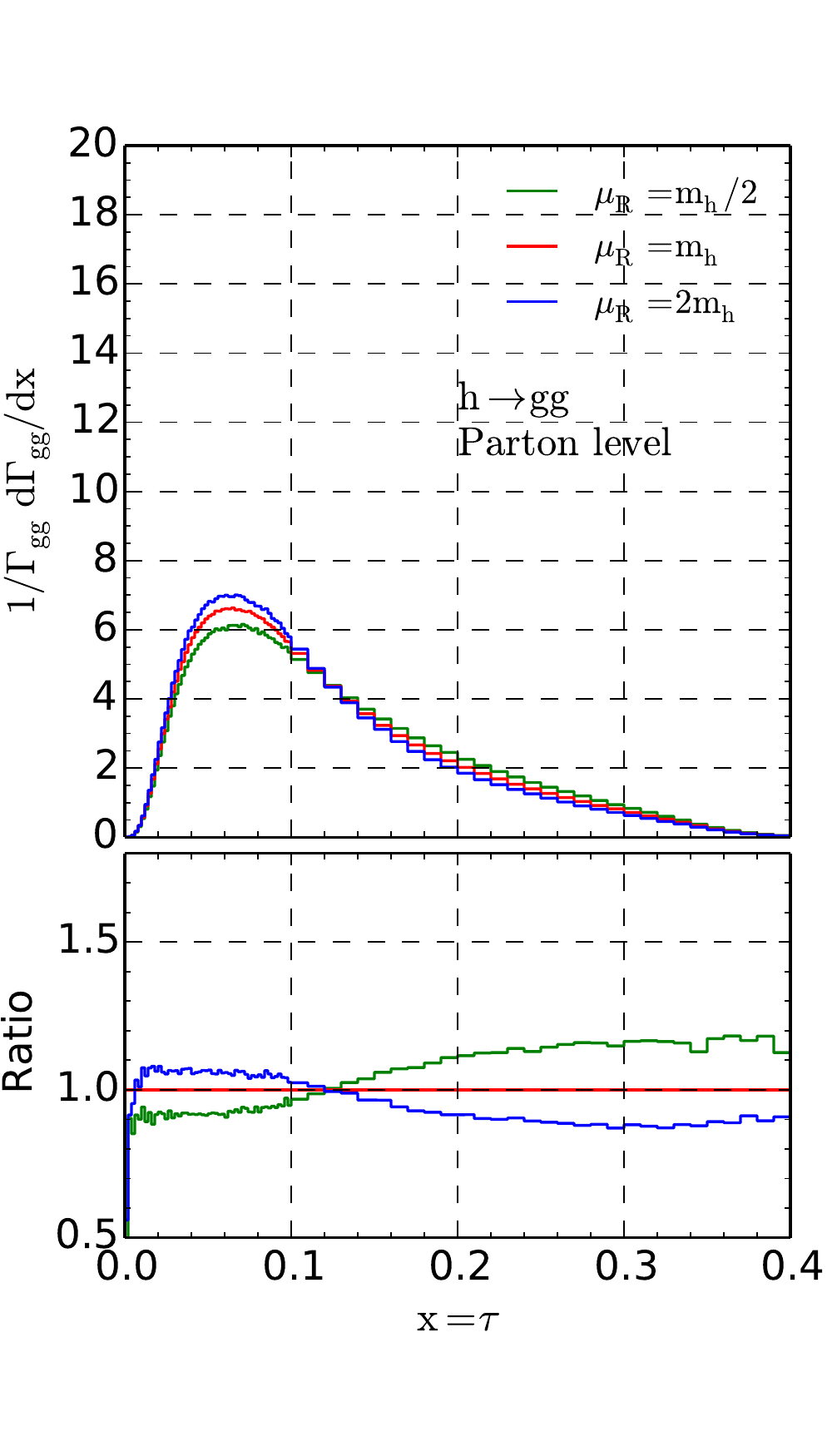}
\includegraphics[width=0.32\textwidth]{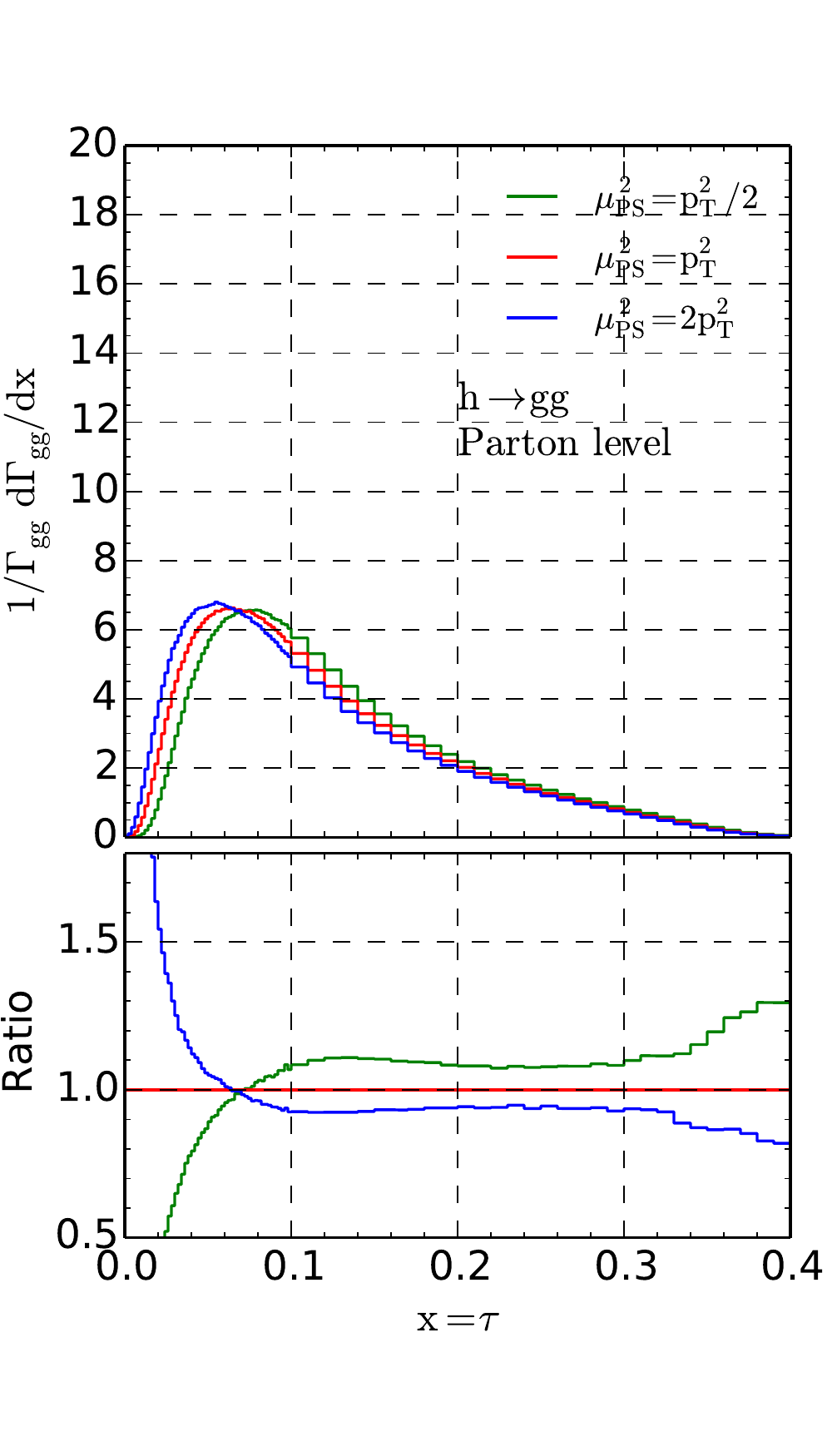}
\caption{
Similar to Fig.~\ref{fig:sysbb} for $\tau$ distributions for the Higgs boson decaying
into gluons. 
\label{fig:sysgg}}
\end{figure}

Lastly we show our matched predictions for the Higgs boson decaying into gluons
in Fig.~\ref{fig:sysgg}.
The $\tau$ distribution peaks at a larger value and has much broader shape
due to stronger radiation strength of gluons.
Thus the normalization of the three-jet sample is relatively larger than in
the case of quarks since we use the same merging scales in the gluon channel
and the quark channel by default.
The size of variations due to the merging scale is within 10\% but is
extended to a broader region comparing to the quark channels.
The variations on renormalization scale are about 10$\sim$20\% for large
$\tau$ values.
The region dominated by the two-jet sample also show a larger dependence on
the renormalization scale comparing to the quark channels in order to balance
the variation of normalizations of the three-jet sample.
The distribution shows a much larger dependence on the scale in parton
shower as expected.
The variations can reach 10\% even for $\tau$ around 0.2 indicating
possible large contributions from missing higher-order matrix elements.
To conclude we have used variations of three scales, namely the merging scale,
renormalization scale, and shower scale to estimate the perturbative
uncertainties of our matched predictions.
We find the variations well cover the full kinematic region with each
scale dominates in its designed region, i.e., the shower scale for small-$\tau$ region,
the merging scale for transition region, and the renormalization scale for
large-$\tau$ region.
In the following we will design our matched calculation with the default
choices of the three scales as the nominal prediction and take the quadratic
sum of variations induced by the three scales as the estimate of full
perturbative uncertainties.
As mentioned earlier our matched calculations are fully exclusive though we
have chosen $\tau$ as the principal variable for merging.
We have checked explicitly our predictions for distributions of the total
hemisphere broadening or the heavy hemisphere mass and found similar
behavior for dependence of the distributions on the three scales.
Our prediction has the same logarithmic accuracy as the usual parton shower
MC in the Sudakov region while it maintains the NLO matched with parton shower
accuracy in the region with three hard jets and the NNLO accuracy for the total partial
width.
Comparisons of our parton-level predictions with results of analytic
resummations are included in Appendix~\ref{sec:appc} that indicates
the logarithmic accuracies of shower MCs are well preserved.

\subsection{Predictions at hadron level}
\begin{figure}[ht]
\centering
\includegraphics[width=0.32\textwidth]{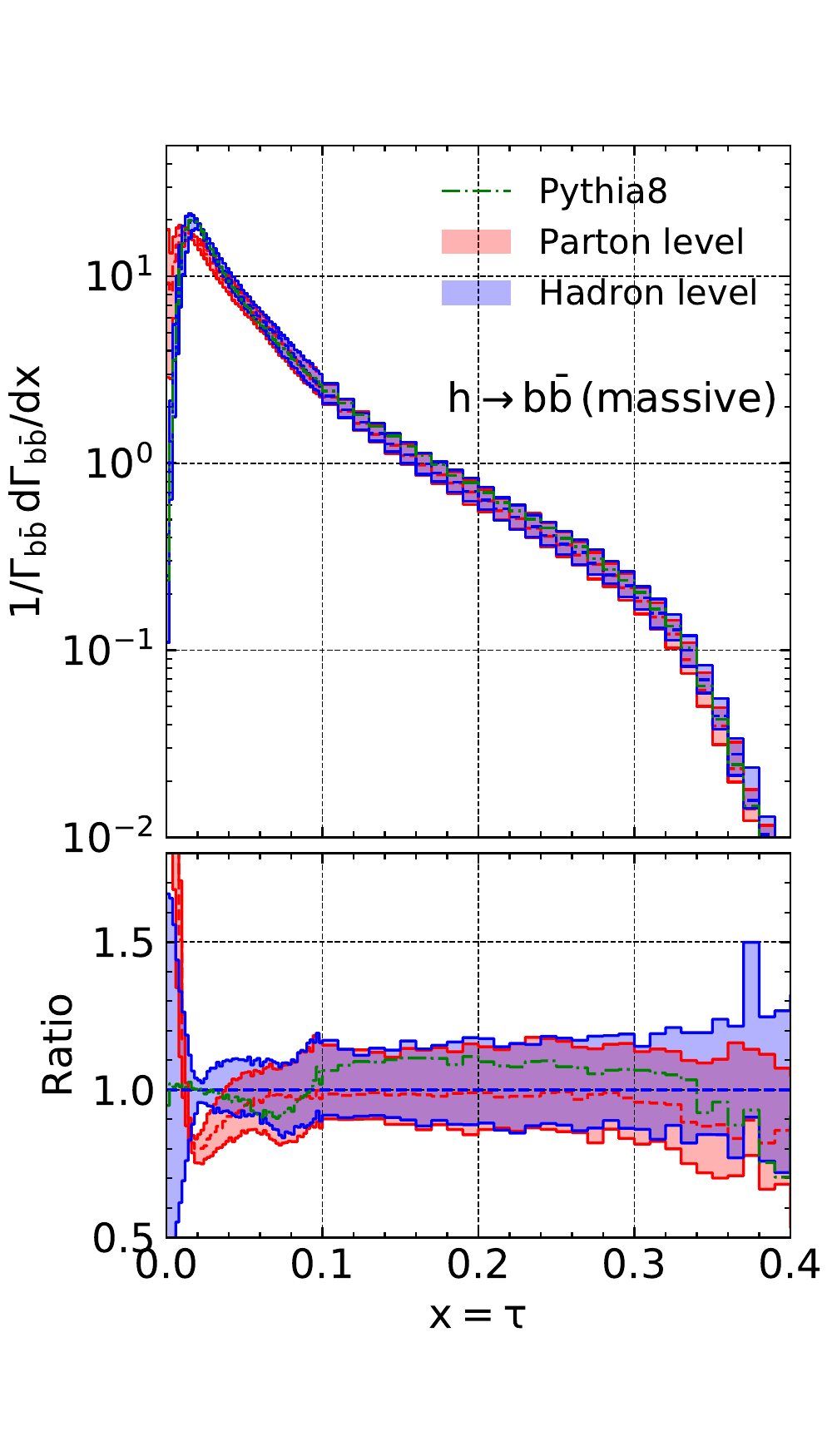}
\includegraphics[width=0.32\textwidth]{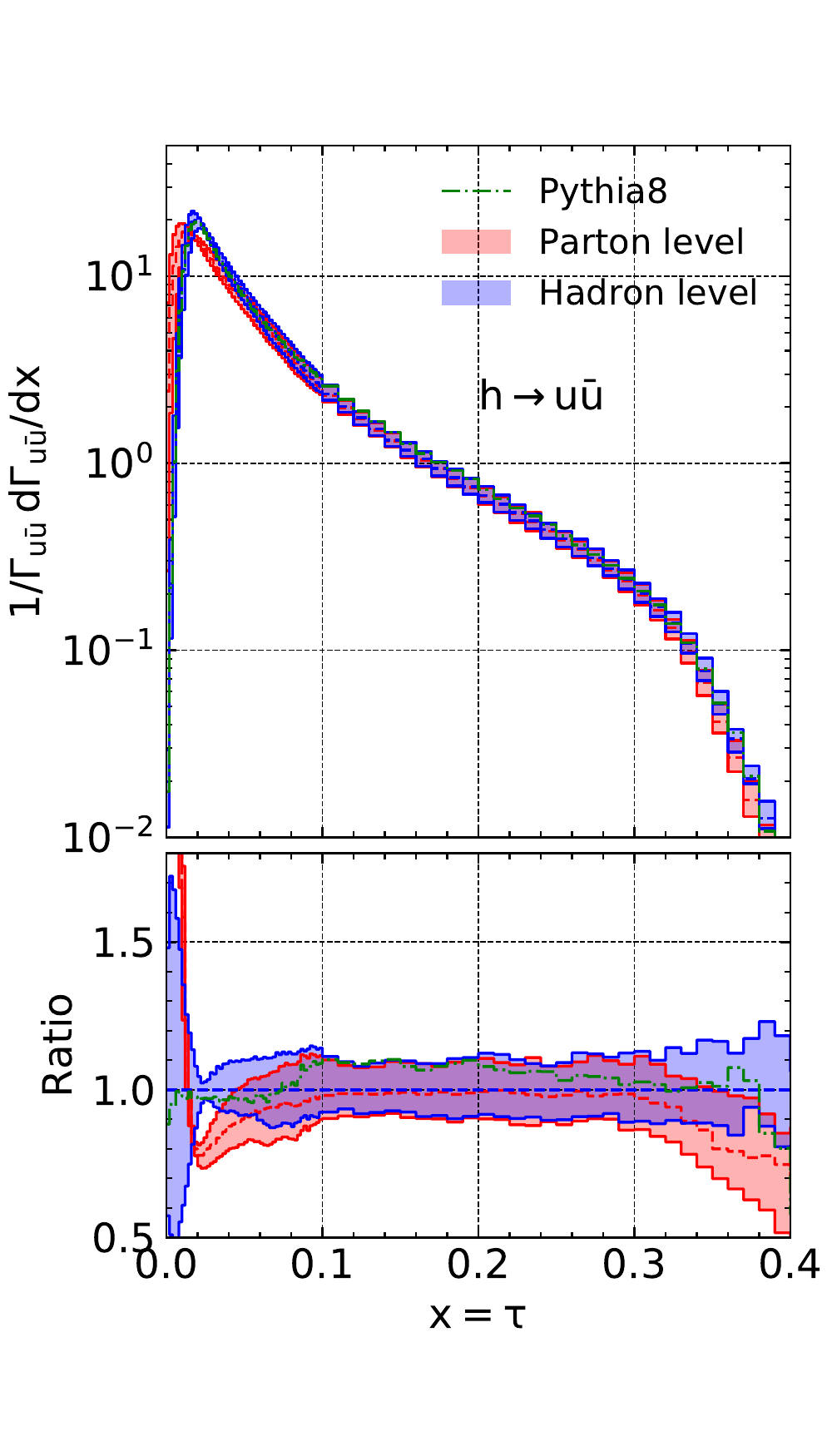}
\includegraphics[width=0.32\textwidth]{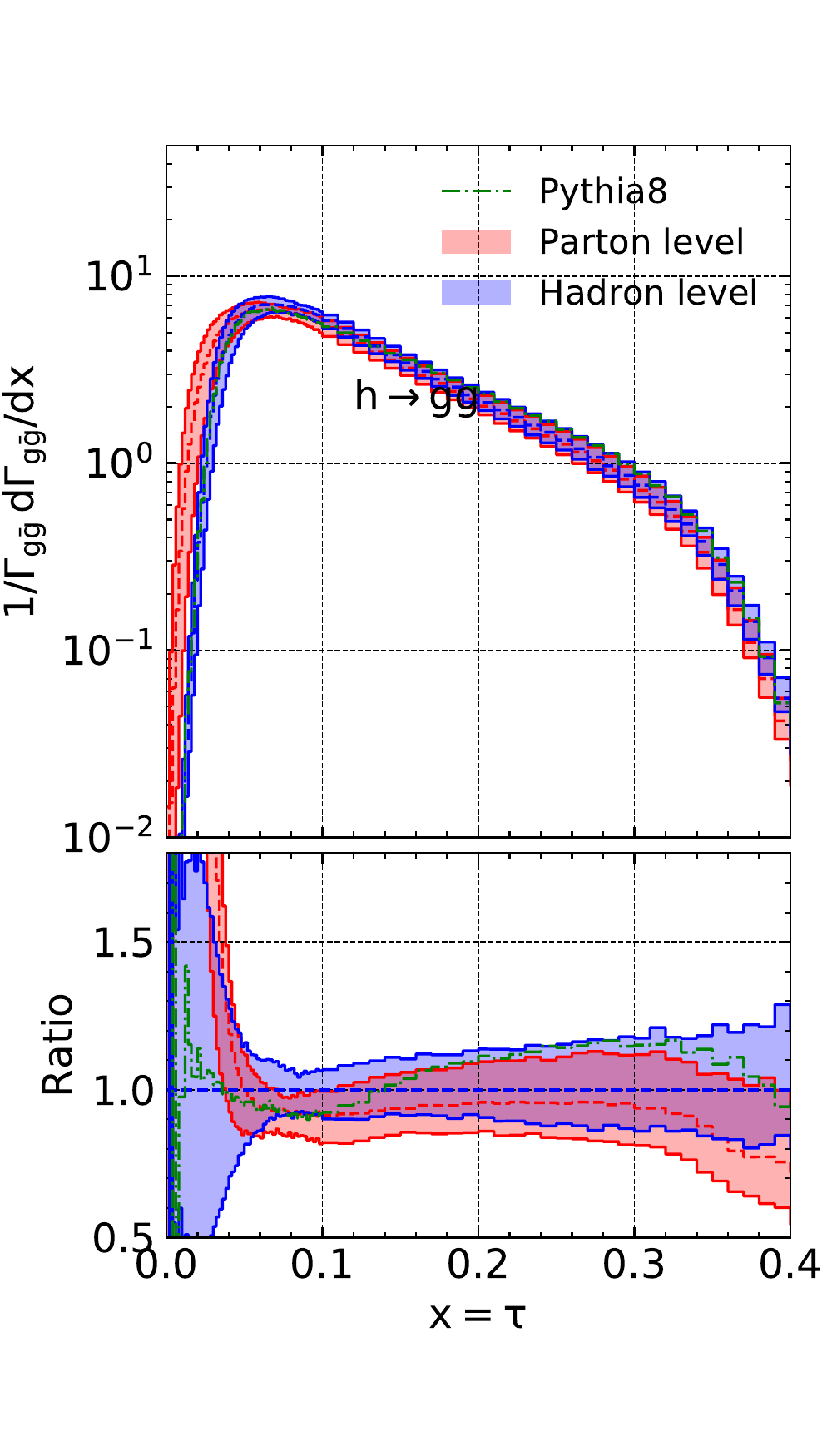}
\caption{
$\tau$ distribution calculated at parton level matched with parton shower,
and further including hadronizations.
The three plots are for the Higgs boson decaying into bottom quarks, up quarks
and gluons respectively.
The colored band indicates the full perturbative uncertainties by adding
variations due to the merging scale, renormalization scale and shower scale
in quadrature.
Predictions from PYTHIA8 at hadron level are also included for comparison.
In each plot the upper panel shows the normalized distributions and the
lower panel shows ratios of different predictions.
\label{fig:fulltau}}
\end{figure}
In the final step we further include hadronization for our matched predictions
which is mandatory for any phenomenological study.
We use the default hadronization model in PYTHIA8.2~\cite{Sjostrand:2014zea} and the Monash tunes~\cite{Skands:2014pea}.
We show the normalized $\tau$ distributions for the Higgs boson decaying into
bottom quarks, up quarks and gluons in Fig.~\ref{fig:fulltau}.
We compare the nominal predictions at parton level and at hadron level with
colored bands indicate the full perturbative uncertainties.
Predictions from PYTHIA8 at hadron level are also included for comparison.
In each plot the upper panel shows the normalized distributions and the
lower panel shows ratios with respect to the nominal prediction at hadron level.
\begin{figure}[ht]
\centering
\includegraphics[width=0.32\textwidth]{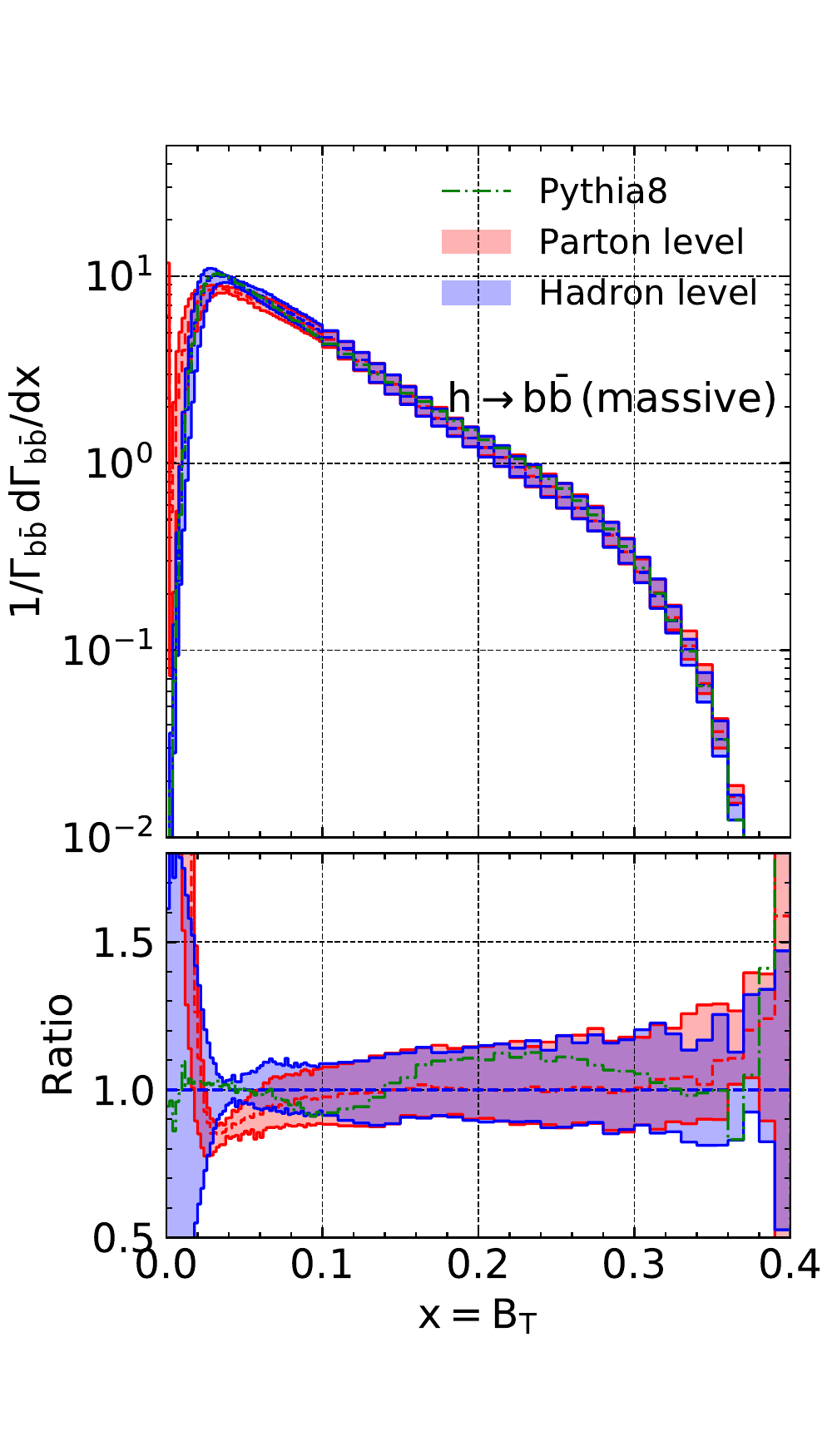}
\includegraphics[width=0.32\textwidth]{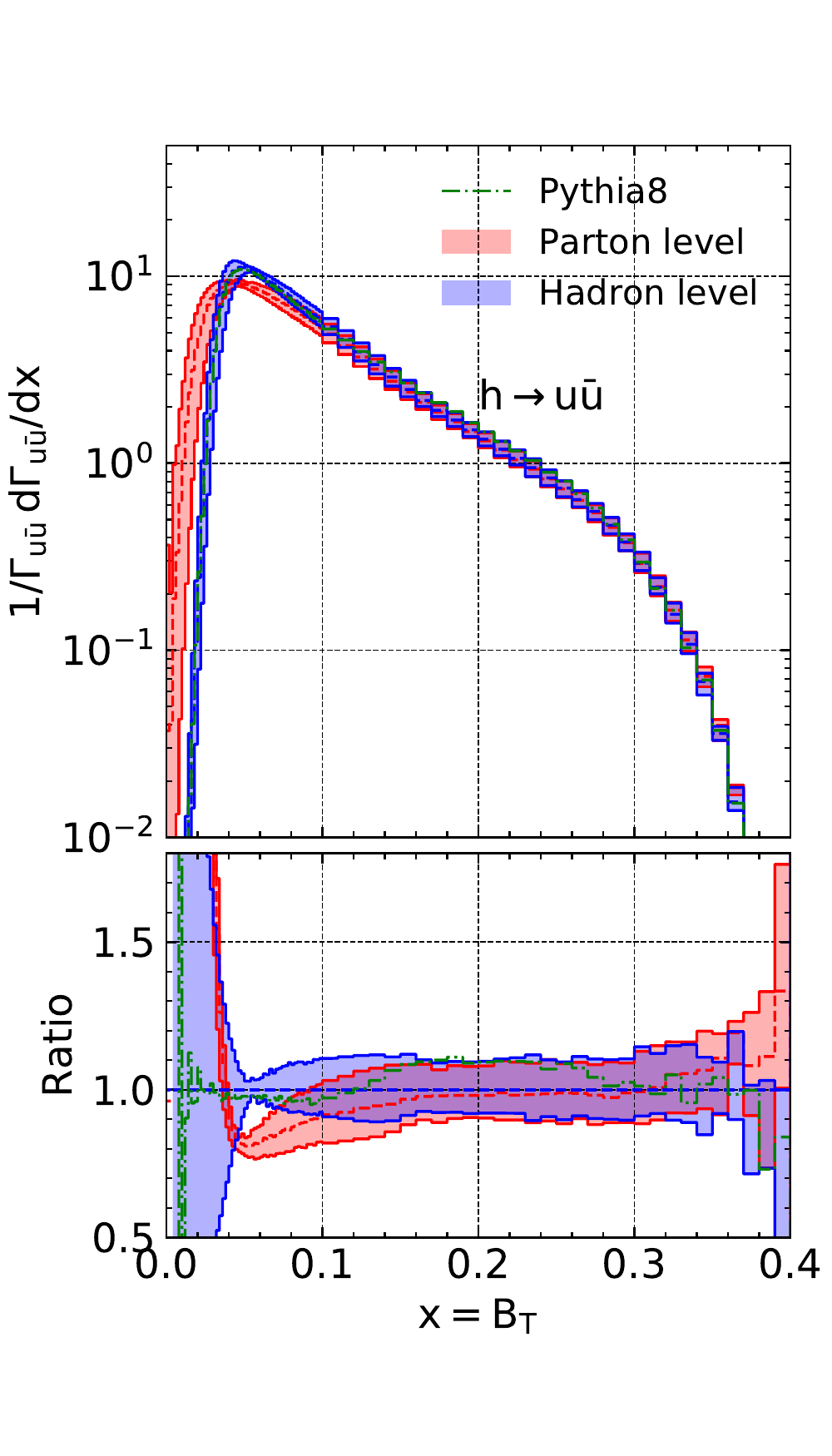}
\includegraphics[width=0.32\textwidth]{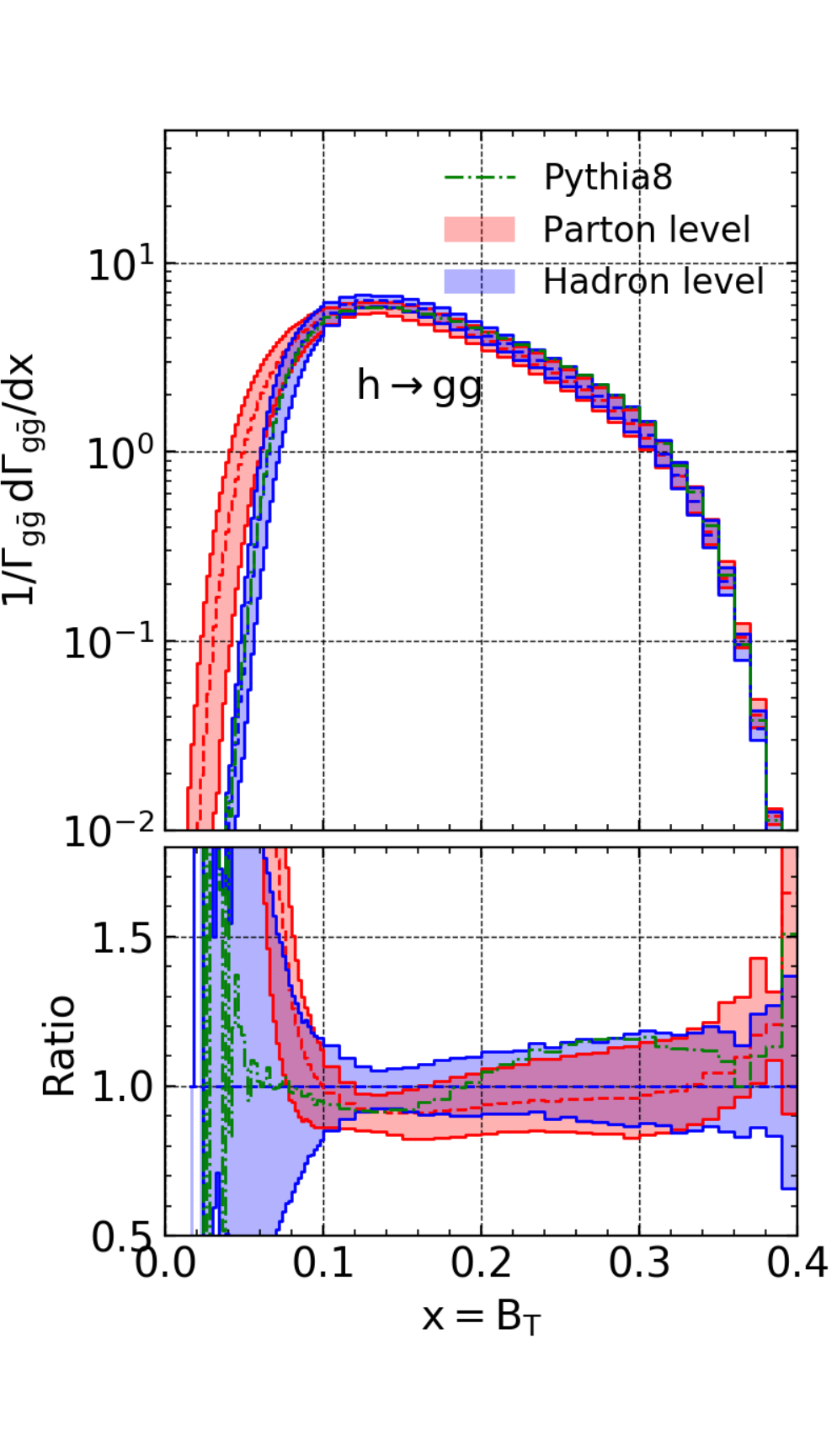}
\caption{
Similar to Fig.~\ref{fig:fulltau} for the distributions of the total hemisphere
broadening.
\label{fig:fullbrod}}
\end{figure}
We found the hadronization corrections
can be as large as tens
percents when close to the peak region and
shift the Sudakov peak to larger $\tau$ values which are well known. 
In the tail region of $\tau$ where the distribution falls off rapidly,
the hadronization and decay of hadrons tend to push the parton-level kinematics
to larger $\tau$ values and thus increase the distribution significantly.
In between the peak and tail region the hadronization corrections are at the
level of 1\% for the quark channels and 5\% for the gluon channel.
The hadronization corrections are almost the same for the decays into bottom
quarks and into up quarks with corrections to the former being slightly smaller due to
the presence of the bottom quark mass.
The size of the perturbative uncertainties are not affected by hadronization.
They range between 10\% to 30\% from the peak region to the tail region, and are even
larger when $\tau$ is to the left of the peak region.
It is interesting that the native hadron-level predictions from
PYTHIA8 lie within our uncertainty bands for the full kinematic region
considered and for all three decay channels.
However they show a harder spectrum in general comparing to our nominal
predictions.
In Fig.~\ref{fig:fullbrod} we show similar results for distributions of
the total hemisphere broadening $B_{T}$.
The distribution peaks at a value of $B_{T}$ that is almost twice of the value of
the $\tau$ distribution for both the quark channel and the gluon channel due to
different dependence on soft and collinear emissions.
In the tail region the $B_T$ distribution falls even more rapidly than $\tau$.
The overall picture on distribution of the total hemisphere broadening is very
similar to that of the $\tau$ distribution including for the size of perturbative
uncertainties and the size of the hadronization corrections.
One difference is that the hadronization corrections tend to decrease the
distribution in the tail region which is opposite to the case of the $\tau$
distribution.
\begin{figure}[ht]
\centering
\includegraphics[width=0.48\textwidth]{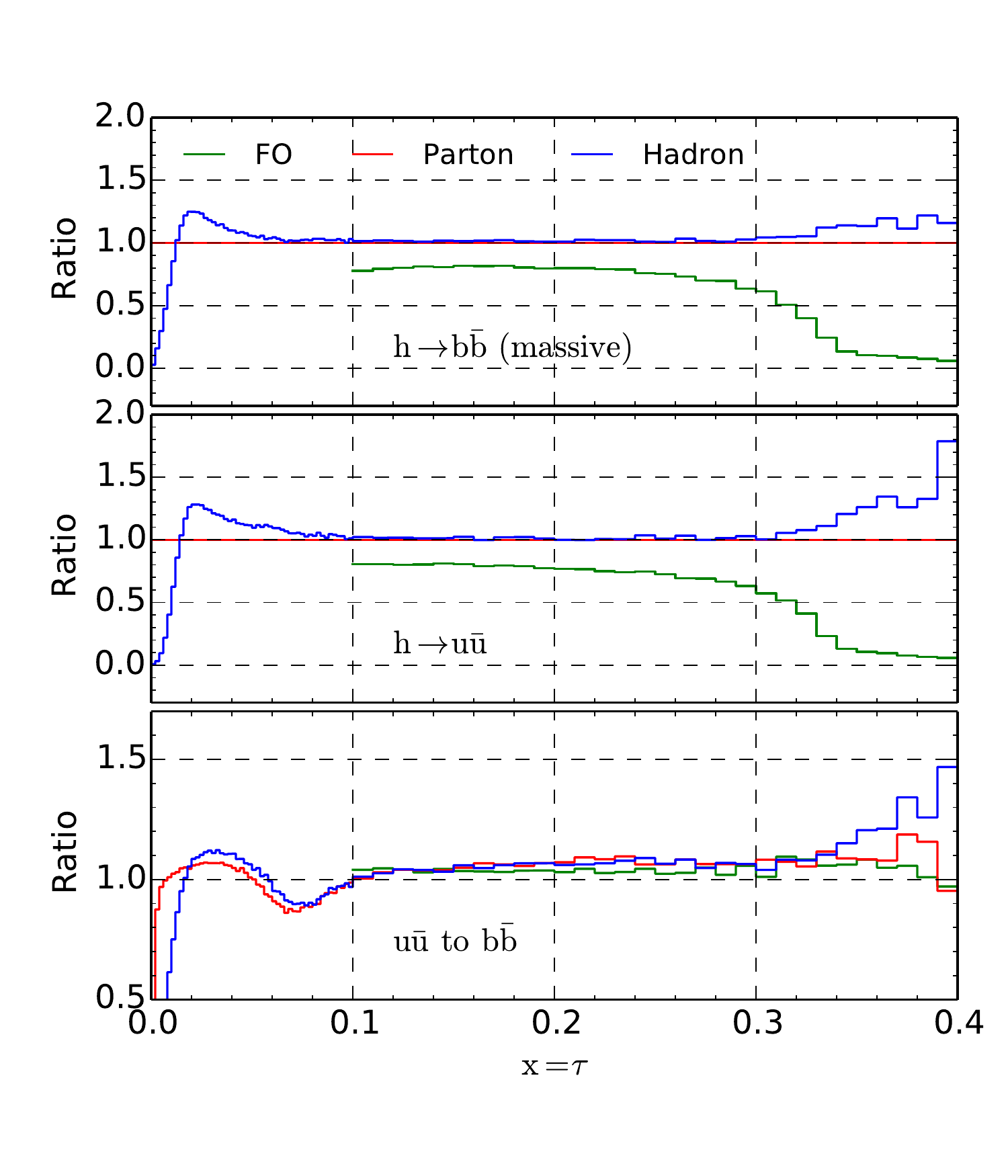}
\caption{
Comparison of the normalized $\tau$ distributions calculated at fixed-order, matching with parton
shower, and further including hadronizations, for the Higgs boson decaying to
bottom quarks (top plot), to up quarks (middle plot).
The fixed-order predictions have been truncated at $\tau=0.1$.
The lower plot shows ratios between the Higgs decaying to up quarks and to bottom quarks
for the corresponding predictions.
\label{fig:focom}}
\end{figure}
In Fig.~\ref{fig:focom} we present a comparison of the matched predictions on
the normalized $\tau$ distribution to the fixed-order predictions (NNLO for the total partial width or NLO for
the distribution) for the decays to bottom quarks and to up quarks respectively.
The fixed-order predictions have been truncated at $\tau=0.1$.
By checking on the ratios of the fixed-order predictions to the matched predictions
at parton level it is evident that by matching to parton shower the
distributions are enhanced by about 20\% for $\tau$ between 0.1 and 0.2, and 40\% for $\tau\sim 0.3$.
That is in agreement with the findings in Ref.~\cite{1901.02253} that higher-order
logarithmic corrections enhance the distributions for massless quarks by a similar
amount.
The fixed-order predictions fail completely at close to the three-jet kinematic
limit of $\tau=1/3$ where multi-parton emissions play an essential role.
In the lower panel of Fig.~\ref{fig:focom} we plot ratios of the
distributions for up quarks to that for bottom quarks with different
calculations.
We observe a suppression of 5\%$\sim$10\% in the distribution of bottom quarks for
$\tau\gtrsim 0.1$ for
both the fixed-order calculation and the two matched calculations at parton
and hadron level.
On the other hand the distribution of bottom quarks peaks at a slightly smaller
$\tau$ value than that of up quarks.
In the merging region it shows the distribution of bottom quarks is 10\% higher
than that of up quarks which could be an artifact due to merging uncertainties.

\section{Summary}\label{sec:con}

In summary we have presented calculations of hadronic decays of the Higgs boson
at NNLO matched with parton shower, including the decays to bottom quarks, light quarks
and gluons.
We use phase space slicing methods to reproduce the NNLO partial widths.
The matched calculations are based on the POWHEG framework together with
PYTHIA8.
We keep the full dependence on the bottom quark mass in the matrix elements thus
ensuring a consistent matching with parton shower.
The calculations provide descriptions on exclusive decays of the Higgs boson
at leading logarithmic accuracy in the Sudakov region and NLO matched with parton shower
accuracy in the three-jet region, with normalizations fixed to the
NNLO partial width.
As examples we show detailed results on distributions of event shape variables
like thrust and total jet broadening.
We found parton shower plays important roles even in the three-jet region where
it increases the NLO distributions by 20\% in general and even more when close to the boundary
of the phase space of fixed orders.
By comparing distributions for bottom quarks and light quarks we found the
normalized distributions are suppressed by 5 to 10\% in the three-jet region
for bottom quarks.
Effects of hadronizations can be significant in the Sudakov region or far tail
region, and are slightly different for bottom quarks and light quarks.
We estimate the total perturbative uncertainties on the distributions and
found they are more than 10\% in the bulk region of the distributions especially
for the decay to gluons.
Predictions at such an accuracy will not be sufficient for measurements 
at future Higgs factories.
However, they can certainly help with evaluations of the physics performances
of the detector designs which are under investigation.  
On another hand, with recent progresses on developments of parton shower
MCs~\cite{2011.10054,2011.04773,2011.04777,2011.15087,2002.11114,2003.06400,2003.00702}
and given the available matrix elements at higher orders~\cite{Mondini:2019vub}, we expect
further improvements on modeling of hadronic decays of the Higgs boson in the near future.

\section*{Acknowledgments}

This work is sponsored by the National Natural
Science Foundation of China under the Grant No. 11875189 and No.11835005.
The authors would like to thank Li Lin Yang, Hua Xing Zhu and Manqi Ruan
for useful discussions.

\appendix

\section{Soft and hard functions at one-loop}\label{sec:appb}
For QCD radiations off two massive quarks the soft function can be calculated using the eikonal
approximation and can be expanded order-by-order in $\alpha_S$.
The Laplace transformed soft function at one-loop can be expressed as~\cite{1408.5134}
\begin{align}
  \tilde{S}(z,L)&\equiv \int_0^{\infty}d \omega 
  \exp{\Big(-\frac{2\omega}{e^{\gamma_E}\kappa}\Big)}S(\omega,z)\nonumber \\
    &=1+\frac{\alpha_S(\mu)}{4\pi}\Big(L\gamma_0^s(z)+c_1(z)\Big)+\mathcal{O}(\alpha_S^2(\mu)),
\end{align}
where $L=\ln(\kappa/\mu)$, and $\gamma_E$ is the Euler-Mascheroni constant.
We have replaced the bottom quark mass with the dimensionless variable,
\begin{equation}
  z=\left(1-\sqrt{1-{4m_b^2}/{m_H^2}}\right)/
  \left(1+\sqrt{1-{4m_b^2}/{m_H^2}}\right).
\end{equation}
The cusp anomalous dimension is given by~\cite{Korchemsky:1987wg,0903.2561,0904.1021},
\begin{equation}
  \gamma_0^s(z)=-8C_F\Big[1+\frac{1+z^2}{1-z^2}\ln(z)\Big],
\end{equation}
and the constant piece of the soft function is
\begin{align}
  c_1(z)&=C_F\Big[\frac{1+z^2}{1-z^2}\big(-2\ln^2(z)+8\ln(z)\ln(1-z)\nonumber\\
   &+8{\rm Li}_2(z)-\frac{4\pi^2}{3}\big)
  -4\frac{1+z}{1-z}\ln(z)\Big].
\end{align}
The hard function can be obtained from the form factor of the $Hb\bar b$
vertex $\hat F$ renormalized in the $\overline {\rm MS}$ scheme
\begin{equation}
  H(Q^2,z)=|\lim_{\epsilon\rightarrow 0} \big[Z_H\hat F(Q^2, z)\big]|^2,
\end{equation}
where $Z_H$ removes fully the remaining infrared divergences in the form
factor.
At one-loop the hard function can be expressed as
\begin{align}
  H(Q^2,z)&=1+\frac{\alpha_S(\mu)}{4\pi}H^{(1)}(Q^2,z)+\mathcal{O}(\alpha^2_S(\mu)),
\end{align}
with 
\begin{align}
  H^{(1)}(Q^2,z)&=\frac{4C_F}{z^2-1}\Big[(z^2+1)(-2{\rm Li}_2(z)-2\ln (z)\ln (1-z)+\frac{\ln^2(z)}{2}
  -\frac{2\pi^2}{3}-1) \nonumber \\
  &+4z\ln(z)+2
  -\ln(\frac{m_b^2}{\mu^2})(1-z^2+(1+z^2)\ln(z))\Big]\nonumber\\
  &+8C_F\Big[1-\frac{3}{4}\ln(\frac{m_b^2}{\mu^2})\Big],
\end{align}
where the renormalization scale dependence in the second line cancels with that from
the soft function and the third line is due to conversion from the on-shell scheme
to the $\overline{\rm MS}$ scheme for the bottom-quark Yukawa coupling.
The two-loop hard function can be obtained in a similar way with the two-loop
form factor calculated in~\cite{hep-ph/0508254,1712.09889}.

\section{Damping factor}\label{sec:appa}
In this appendix we include further details on the damped predictions used.
We can rewrite it as
\begin{align}
  \Gamma^{damp.}&=\int dx\mathcal{D}(x)\Big(\frac{d\Gamma_s(x)}{dx}
  -D^{(1)}(x)\Gamma_0\frac{d\Gamma^{(1)}_{s}(x)}{dx}\Big) \nonumber\\
   &+\int dx\mathcal{D}(x)\Big(\Big[\frac{d\Gamma_{3j}(x)}{dx}-\frac{d\Gamma_s(x)}{dx}\Big]
   -D^{(1)}(x)\Gamma_0\Big[\frac{d\Gamma^{(1)}_{3j}(x)}{dx}-\frac{d\Gamma^{(1)}_s(x)}{dx}\Big]\Big),
\end{align}
where we have simply added and subtracted a same term from the original definition
in Eq.~(\ref{eq:damp}).
Now the integrand of the first line becomes a total derivative.
In the second line each piece in the brackets is now regularized for $x$ goes to
zero and the damping factor in front can be expanded order-by-order.
To be specific, we find
\begin{align}\label{eq:mf}
  \Gamma^{damp.}&=\Gamma_0\int dx \frac{d\mathcal{D}(x)}{dx}
  +\int dx\Big[\frac{d\Gamma_{3j}(x)}{dx}-\frac{d\Gamma_s(x)}{dx}\Big]
  +\mathcal{O}(\alpha^3_S(\mu)), \nonumber\\
  &=\Gamma^{NNLO}+\mathcal{O}(\alpha^3_S(\mu)).
\end{align}
In deriving the last step one can imagine adding an arbitrarily small cutoff
as the lower limit of the integration, and the formula simply goes back
to Eq.~(\ref{eq:fo}) for the phase-space slicing method.
In all our derivations we have fixed the renormalization scale rather
than using a dynamic scale varying with $x$.
Before passing the reweighted three-jet matrix elements to POWHEG we would
like to further eliminate those spurious $\mathcal{O}(\alpha_S^3)$ terms in
Eq.~(\ref{eq:mf}).
In addition we require distributions in the resolved three-jet region to
be less modified by the reweighting.
These can be achieved by using a slightly modified damping factor as below,
\begin{align}
  \tilde{D}^{(1,2)}(x)\equiv  D^{(1,2)}(x')\frac{1}{1+(x/x_0)^4},\,\,\,
  \mathcal{\tilde{D}}(x)\equiv\exp{[\tilde{D}^{(1)}(x)+\tilde{D}^{(2)}(x)+d^{(3)}(x_0,\mu)]},
\end{align}
where $x_0$ is a variable to control the functional range of the reweighting,
and $x'$ is defined as
\begin{equation}
  x'=\left(1+\frac{1}{{\rm min}(x,x_0)}-\frac{1}{x_0}\right)^{-1}.
\end{equation}
$d^{(3)}(x_0, \mu)$ is a constant term tuned numerically to eliminate spurious terms
of $\mathcal{O}(\alpha_S^3)$.
It depends on the preselected value of $x_0$ as well as on the renormalization scale.

\section{Comparison to analytic resummations}\label{sec:appc}
\begin{figure}[ht]
\centering
\includegraphics[width=0.48\textwidth]{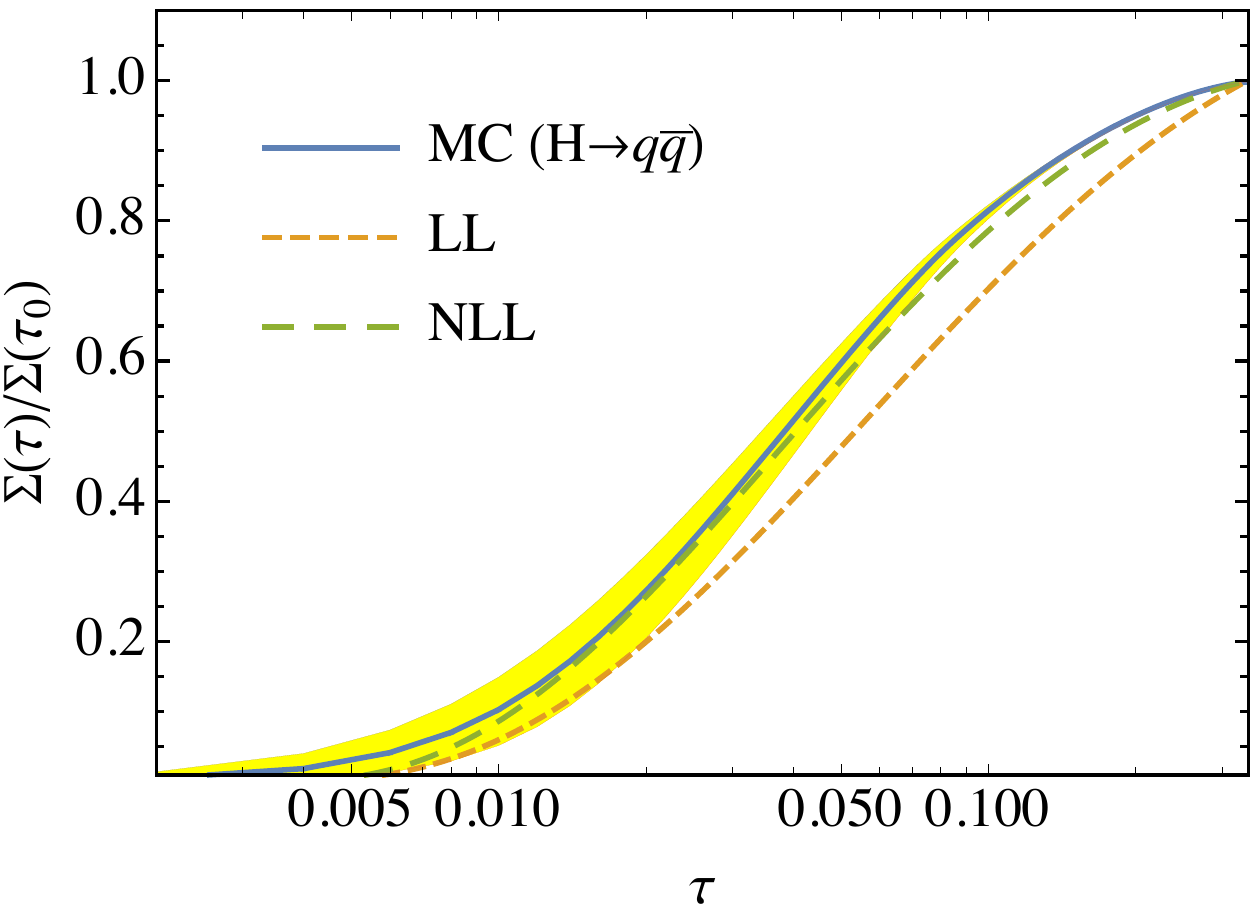}\hspace{0.1in}
\includegraphics[width=0.48\textwidth]{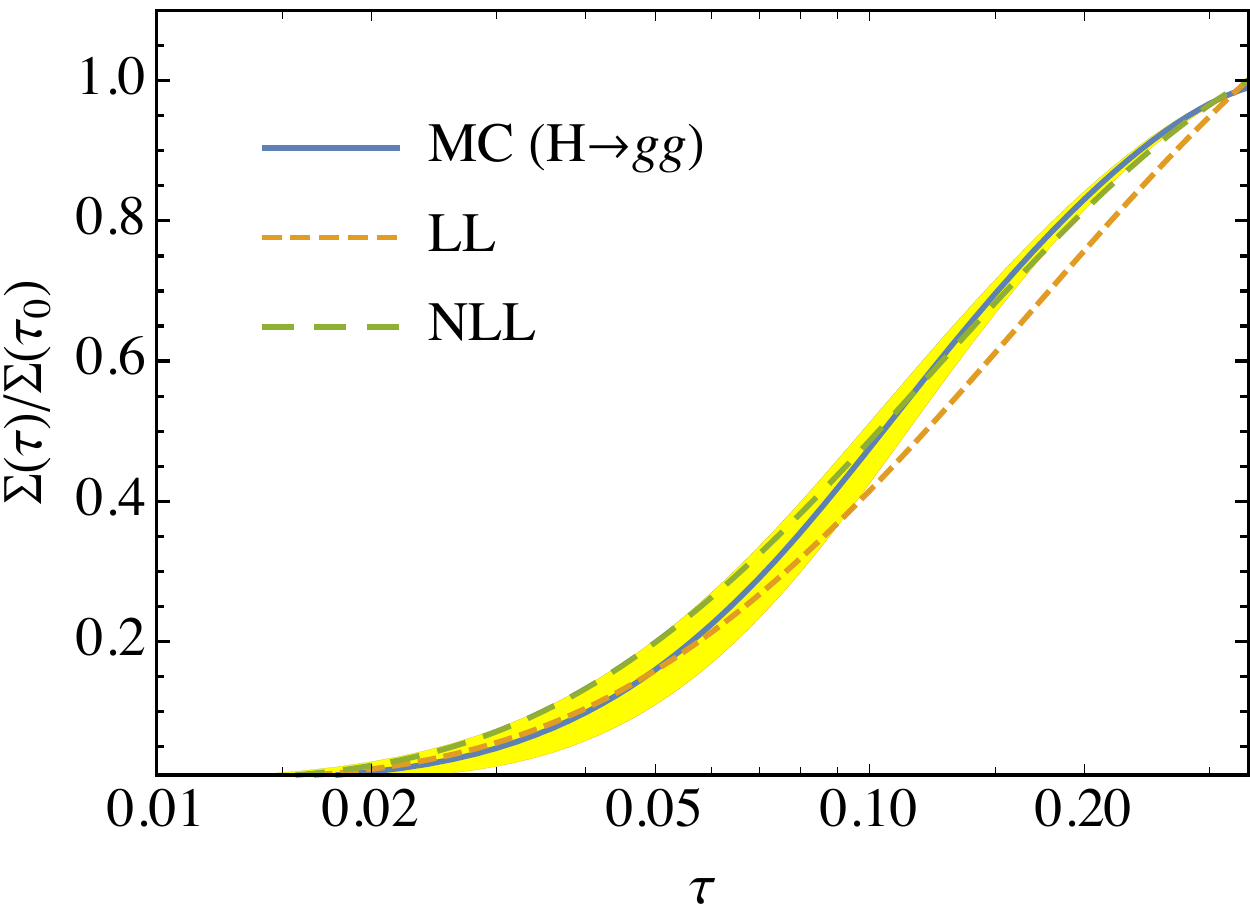}
\caption{
Comparison of predictions on the cumulate width as a function of $\tau$,
as from the parton-level MCs and the analytic resummations at LL and NLL
accuracy.
The colored bands indicate variations due to the shower scale.
All predictions have been normalized to the respective value at the
3-jet kinematic limit.
\label{fig:recom1}}
\end{figure}

\begin{figure}[ht]
\centering
\includegraphics[width=0.48\textwidth]{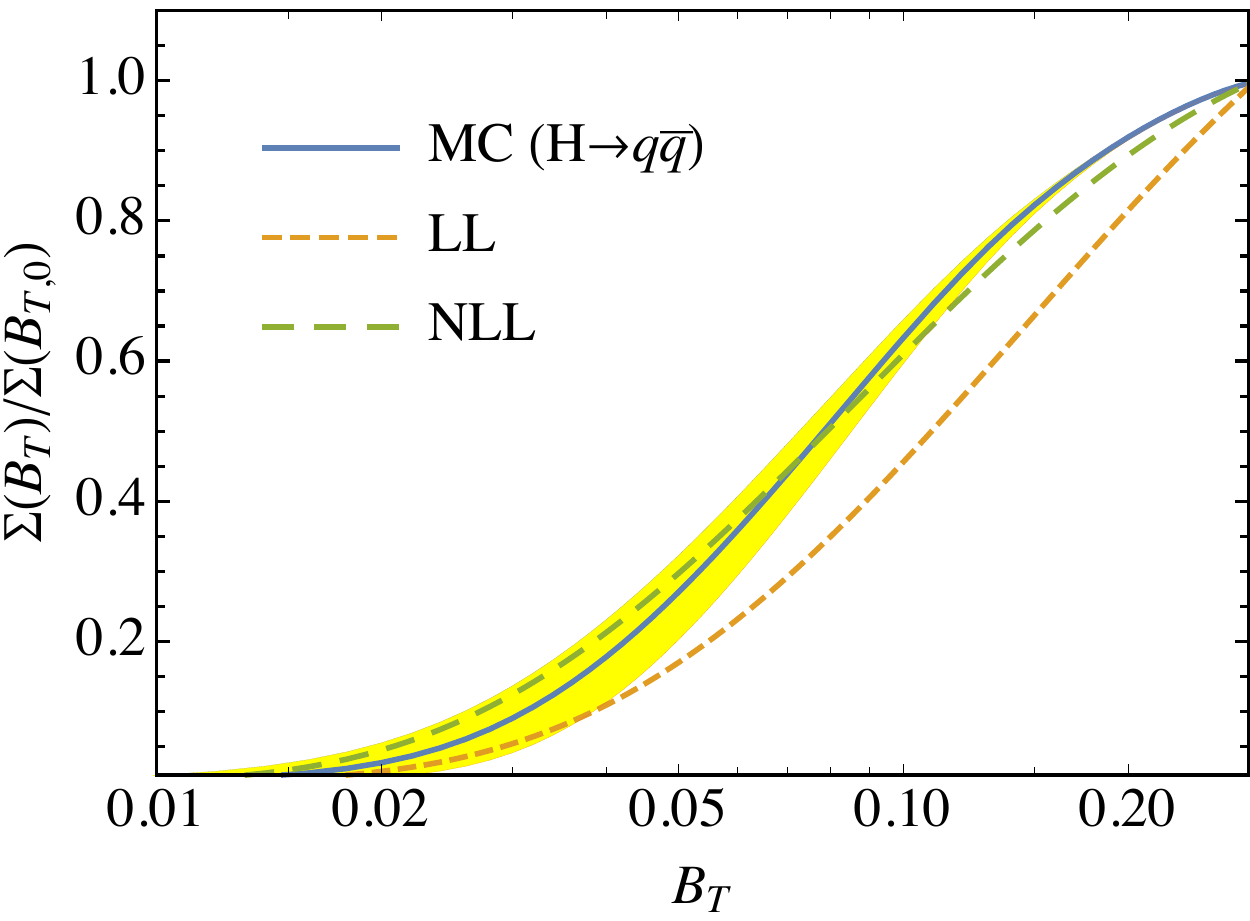}\hspace{0.1in}
\includegraphics[width=0.48\textwidth]{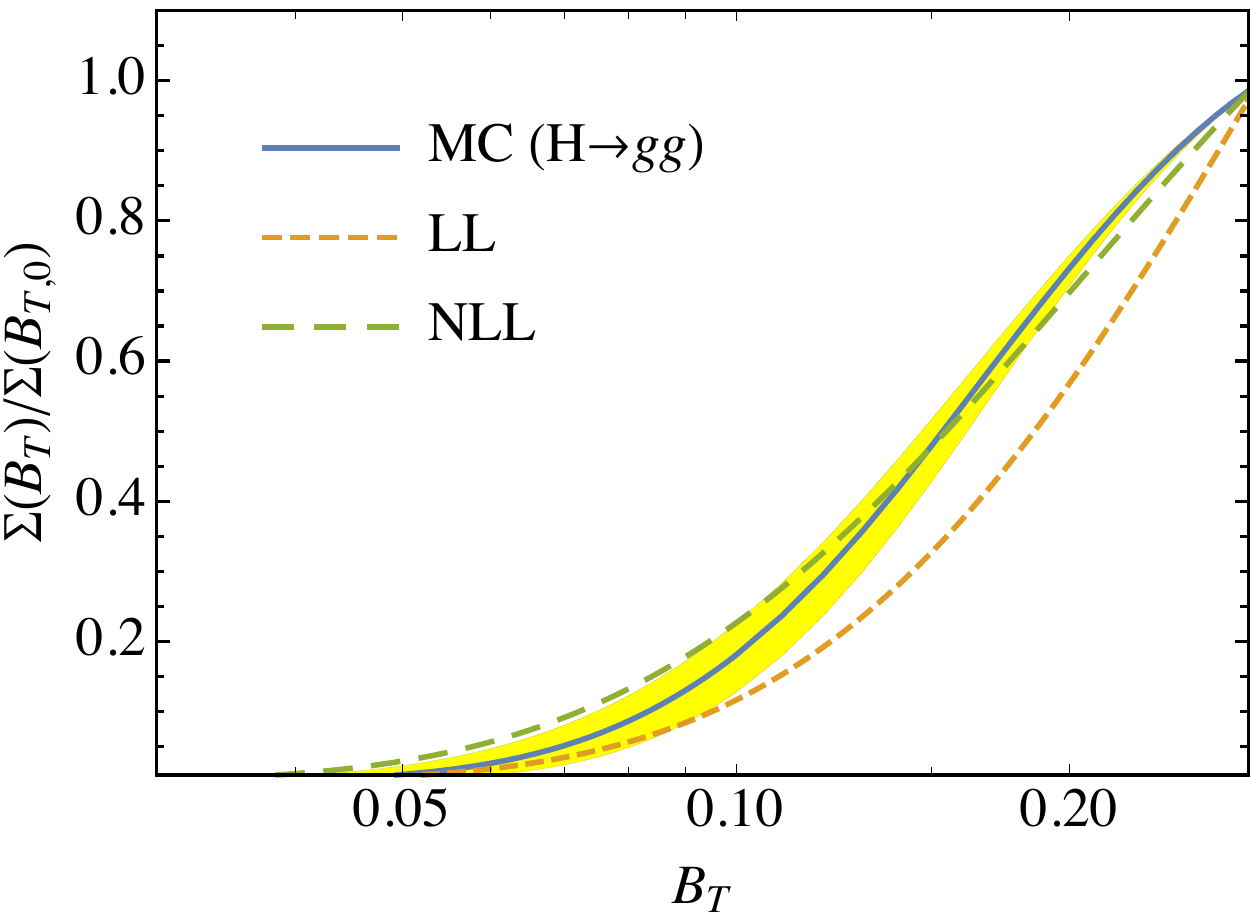}
\caption{
Comparison of predictions on the cumulate width as a function of $B_T$,
as from the parton-level MCs and the analytic resummations at LL and NLL
accuracy.
The color bands indicate variations due to the shower scale.
All predictions have been normalized to the respective value at the
3-jet kinematic limit.
\label{fig:recom2}}
\end{figure}

We compare our predictions for the Higgs boson decaying into massless quarks or
gluons at the parton level to results from analytic resummations.
The factorization formulas and renormalization group evolutions can be found in
Ref.~\cite{0803.0342} for thrust and~\cite{Becher:2011pf} for total hemisphere broadening.
We do not reproduce details of the analytic resummations but only show numerical
results at leading logarithmic (LL) and next-to-leading logarithmic (NLL) accuracy
with a choice of $\alpha_S(M_Z)=0.118$.
In Fig.~\ref{fig:recom1} we show the cumulate width as a function of $\tau$
normalized to its value at the 3-jet kinematic limit, $\tau_0=1/3$, for
the Higgs boson decaying into massless quarks and gluons respectively.
The colored bands indicate the variations when changing the square of the shower
scale by a factor of two.
We find that when close to the Sudakov region, the MC predictions in general
lie between the LL and NLL results, and are more closer to the NLL ones.
Fig.~\ref{fig:recom2} shows similar comparisons for $B_T$.
We again find good agreements between the MC predictions and the analytic resummations.
We conclude that the usual logarithmic accuracy in parton shower MCs are well
preserved in our matched predictions.

\bibliography{hnnlops}

\providecommand{\href}[2]{#2}\begingroup\raggedright\begin{thebibliography}{100}

\bibitem{Aad:2012tfa}
{\bf ATLAS} Collaboration, G.~Aad et~al., {\it {Observation of a new particle
  in the search for the Standard Model Higgs boson with the ATLAS detector at
  the LHC}},  {\em Phys. Lett. B} {\bf 716} (2012) 1--29,
  [\href{http://arxiv.org/abs/1207.7214}{{\tt arXiv:1207.7214}}].

\bibitem{Chatrchyan:2012xdj}
{\bf CMS} Collaboration, S.~Chatrchyan et~al., {\it {Observation of a New Boson
  at a Mass of 125 GeV with the CMS Experiment at the LHC}},  {\em Phys. Lett.
  B} {\bf 716} (2012) 30--61, [\href{http://arxiv.org/abs/1207.7235}{{\tt
  arXiv:1207.7235}}].

\bibitem{Behnke:2013xla}
{\it {The International Linear Collider Technical Design Report - Volume 1:
  Executive Summary}},  \href{http://arxiv.org/abs/1306.6327}{{\tt
  arXiv:1306.6327}}.

\bibitem{CEPCStudyGroup:2018ghi}
{\bf CEPC Study Group} Collaboration, M.~Dong et~al., {\it {CEPC Conceptual
  Design Report: Volume 2 - Physics \textbackslash{}\& Detector}},
  \href{http://arxiv.org/abs/1811.10545}{{\tt arXiv:1811.10545}}.

\bibitem{Lebrun:2012hj}
P.~Lebrun, L.~Linssen, A.~Lucaci-Timoce, D.~Schulte, F.~Simon, S.~Stapnes,
  N.~Toge, H.~Weerts, and J.~Wells, {\it {The CLIC Programme: Towards a Staged
  e+e- Linear Collider Exploring the Terascale : CLIC Conceptual Design
  Report}},  \href{http://arxiv.org/abs/1209.2543}{{\tt arXiv:1209.2543}}.

\bibitem{Gomez-Ceballos:2013zzn}
{\bf TLEP Design Study Working Group} Collaboration, M.~Bicer et~al., {\it
  {First Look at the Physics Case of TLEP}},  {\em JHEP} {\bf 01} (2014) 164,
  [\href{http://arxiv.org/abs/1308.6176}{{\tt arXiv:1308.6176}}].

\bibitem{1901.06150}
J.~P. Delahaye, M.~Diemoz, K.~Long, B.~Mansouli\'e, N.~Pastrone, L.~Rivkin,
  D.~Schulte, A.~Skrinsky, and A.~Wulzer, {\it {Muon Colliders}},
  \href{http://arxiv.org/abs/1901.06150}{{\tt arXiv:1901.06150}}.

\bibitem{1307.1347}
{\bf LHC Higgs Cross Section Working Group} Collaboration, J.~R. Andersen
  et~al., {\it {Handbook of LHC Higgs Cross Sections: 3. Higgs Properties}},
  \href{http://arxiv.org/abs/1307.1347}{{\tt arXiv:1307.1347}}.

\bibitem{1810.09037}
F.~An et~al., {\it {Precision Higgs physics at the CEPC}},  {\em Chin. Phys. C}
  {\bf 43} (2019), no.~4 043002, [\href{http://arxiv.org/abs/1810.09037}{{\tt
  arXiv:1810.09037}}].

\bibitem{1906.05379}
A.~Freitas et~al., {\it {Theoretical uncertainties for electroweak and
  Higgs-boson precision measurements at FCC-ee}},
  \href{http://arxiv.org/abs/1906.05379}{{\tt arXiv:1906.05379}}.

\bibitem{Davies:2017xsp}
J.~Davies, M.~Steinhauser, and D.~Wellmann, {\it {Completing the hadronic Higgs
  boson decay at order $\alpha_s^4$}},  {\em Nucl. Phys. B} {\bf 920} (2017)
  20--31, [\href{http://arxiv.org/abs/1703.02988}{{\tt arXiv:1703.02988}}].

\bibitem{Baikov:2006ch}
P.~Baikov and K.~Chetyrkin, {\it {Top Quark Mediated Higgs Boson Decay into
  Hadrons to Order $\alpha_s^5$}},  {\em Phys. Rev. Lett.} {\bf 97} (2006)
  061803, [\href{http://arxiv.org/abs/hep-ph/0604194}{{\tt hep-ph/0604194}}].

\bibitem{Herzog:2017dtz}
F.~Herzog, B.~Ruijl, T.~Ueda, J.~Vermaseren, and A.~Vogt, {\it {On Higgs decays
  to hadrons and the R-ratio at N$^{4}$LO}},  {\em JHEP} {\bf 08} (2017) 113,
  [\href{http://arxiv.org/abs/1707.01044}{{\tt arXiv:1707.01044}}].

\bibitem{Denner:2011mq}
A.~Denner, S.~Heinemeyer, I.~Puljak, D.~Rebuzzi, and M.~Spira, {\it {Standard
  Model Higgs-Boson Branching Ratios with Uncertainties}},  {\em Eur. Phys. J.
  C} {\bf 71} (2011) 1753, [\href{http://arxiv.org/abs/1107.5909}{{\tt
  arXiv:1107.5909}}].

\bibitem{Spira:2016ztx}
M.~Spira, {\it {Higgs Boson Production and Decay at Hadron Colliders}},  {\em
  Prog. Part. Nucl. Phys.} {\bf 95} (2017) 98--159,
  [\href{http://arxiv.org/abs/1612.07651}{{\tt arXiv:1612.07651}}].

\bibitem{1905.12903}
Y.~Bai, C.~Chen, Y.~Fang, G.~Li, M.~Ruan, J.-Y. Shi, B.~Wang, P.-Y. Kong, B.-Y.
  Lan, and Z.-F. Liu, {\it {Measurements of decay branching fractions of $H\to
  b\bar{b}/c\bar{c}/gg$ in associated $(e^{+}e^{-}/\mu^{+}\mu^{-})H$ production
  at the CEPC}},  {\em Chin. Phys. C} {\bf 44} (2020), no.~1 013001,
  [\href{http://arxiv.org/abs/1905.12903}{{\tt arXiv:1905.12903}}].

\bibitem{1812.09478}
Y.~Zhu and M.~Ruan, {\it {Performance study of the separation of the full
  hadronic WW and ZZ events at the CEPC}},
  \href{http://arxiv.org/abs/1812.09478}{{\tt arXiv:1812.09478}}.

\bibitem{1608.01746}
J.~Gao, {\it {Probing light-quark Yukawa couplings via hadronic event shapes at
  lepton colliders}},  {\em JHEP} {\bf 01} (2018) 038,
  [\href{http://arxiv.org/abs/1608.01746}{{\tt arXiv:1608.01746}}].

\bibitem{2009.02000}
Q.~Bi, K.~Chai, J.~Gao, Y.~Liu, and H.~Zhang, {\it {Investigating Bottom-Quark
  Yukawa Interaction at Higgs Factory}},
  \href{http://arxiv.org/abs/2009.02000}{{\tt arXiv:2009.02000}}.

\bibitem{0909.1521}
D.~E. Kaplan and M.~McEvoy, {\it {Searching for Higgs decays to four bottom
  quarks at LHCb}},  {\em Phys. Lett. B} {\bf 701} (2011) 70--74,
  [\href{http://arxiv.org/abs/0909.1521}{{\tt arXiv:0909.1521}}].

\bibitem{1312.4992}
D.~Curtin et~al., {\it {Exotic decays of the 125 GeV Higgs boson}},  {\em Phys.
  Rev. D} {\bf 90} (2014), no.~7 075004,
  [\href{http://arxiv.org/abs/1312.4992}{{\tt arXiv:1312.4992}}].

\bibitem{Liu:2016zki}
Z.~Liu, L.-T. Wang, and H.~Zhang, {\it {Exotic decays of the 125 GeV Higgs
  boson at future $e^+e^-$ lepton colliders}},  {\em Chin. Phys. C} {\bf 41}
  (2017), no.~6 063102, [\href{http://arxiv.org/abs/1612.09284}{{\tt
  arXiv:1612.09284}}].

\bibitem{Liu:2016ahc}
S.~Liu, Y.-L. Tang, C.~Zhang, and S.-h. Zhu, {\it {Exotic Higgs Decay
  $h\rightarrow\phi\phi\rightarrow 4b$ at the LHeC}},  {\em Eur. Phys. J. C}
  {\bf 77} (2017), no.~7 457, [\href{http://arxiv.org/abs/1608.08458}{{\tt
  arXiv:1608.08458}}].

\bibitem{1905.04865}
J.~Gao, {\it {Higgs boson decay into four bottom quarks in the SM and beyond}},
   {\em JHEP} {\bf 08} (2019) 174, [\href{http://arxiv.org/abs/1905.04865}{{\tt
  arXiv:1905.04865}}].

\bibitem{Anastasiou:2011qx}
C.~Anastasiou, F.~Herzog, and A.~Lazopoulos, {\it {The fully differential decay
  rate of a Higgs boson to bottom-quarks at NNLO in QCD}},  {\em JHEP} {\bf 03}
  (2012) 035, [\href{http://arxiv.org/abs/1110.2368}{{\tt arXiv:1110.2368}}].

\bibitem{DelDuca:2015zqa}
V.~Del~Duca, C.~Duhr, G.~Somogyi, F.~Tramontano, and Z.~Tr\'ocs\'anyi, {\it
  {Higgs boson decay into b-quarks at NNLO accuracy}},  {\em JHEP} {\bf 04}
  (2015) 036, [\href{http://arxiv.org/abs/1501.07226}{{\tt arXiv:1501.07226}}].

\bibitem{Mondini:2019gid}
R.~Mondini, M.~Schiavi, and C.~Williams, {\it {N$^{3}$LO predictions for the
  decay of the Higgs boson to bottom quarks}},  {\em JHEP} {\bf 06} (2019) 079,
  [\href{http://arxiv.org/abs/1904.08960}{{\tt arXiv:1904.08960}}].

\bibitem{Bernreuther:2018ynm}
W.~Bernreuther, L.~Chen, and Z.-G. Si, {\it {Differential decay rates of
  CP-even and CP-odd Higgs bosons to top and bottom quarks at NNLO QCD}},  {\em
  JHEP} {\bf 07} (2018) 159, [\href{http://arxiv.org/abs/1805.06658}{{\tt
  arXiv:1805.06658}}].

\bibitem{1907.05398}
F.~Caola, K.~Melnikov, and R.~R\"ontsch, {\it {Analytic results for decays of
  color singlets to $gg$ and $q \bar q$ final states at NNLO QCD with the
  nested soft-collinear subtraction scheme}},  {\em Eur. Phys. J. C} {\bf 79}
  (2019), no.~12 1013, [\href{http://arxiv.org/abs/1907.05398}{{\tt
  arXiv:1907.05398}}].

\bibitem{1911.11524}
A.~Behring and W.~Bizo\'n, {\it {Higgs decay into massive b-quarks at NNLO QCD
  in the nested soft-collinear subtraction scheme}},  {\em JHEP} {\bf 01}
  (2020) 189, [\href{http://arxiv.org/abs/1911.11524}{{\tt arXiv:1911.11524}}].

\bibitem{2007.15015}
G.~Somogyi and F.~Tramontano, {\it {Fully exclusive heavy quark-antiquark pair
  production from a colourless initial state at NNLO in QCD}},  {\em JHEP} {\bf
  11} (2020) 142, [\href{http://arxiv.org/abs/2007.15015}{{\tt
  arXiv:2007.15015}}].

\bibitem{Li:2018qiy}
G.~Li, Z.~Li, Y.~Liu, Y.~Wang, and X.~Zhao, {\it {Probing the Higgs boson-gluon
  coupling via the jet energy profile at $e^+e^-$ colliders}},  {\em Phys. Rev.
  D} {\bf 98} (2018), no.~7 076010,
  [\href{http://arxiv.org/abs/1805.10138}{{\tt arXiv:1805.10138}}].

\bibitem{1903.07277}
M.-X. Luo, V.~Shtabovenko, T.-Z. Yang, and H.~X. Zhu, {\it {Analytic
  Next-To-Leading Order Calculation of Energy-Energy Correlation in
  Gluon-Initiated Higgs Decays}},  {\em JHEP} {\bf 06} (2019) 037,
  [\href{http://arxiv.org/abs/1903.07277}{{\tt arXiv:1903.07277}}].

\bibitem{Gao:2020vyx}
J.~Gao, V.~Shtabovenko, and T.-Z. Yang, {\it {Energy-energy correlation in
  hadronic Higgs decays: analytic results and phenomenology at NLO}},
  \href{http://arxiv.org/abs/2012.14188}{{\tt arXiv:2012.14188}}.

\bibitem{1901.02253}
J.~Gao, Y.~Gong, W.-L. Ju, and L.~L. Yang, {\it {Thrust distribution in Higgs
  decays at the next-to-leading order and beyond}},  {\em JHEP} {\bf 03} (2019)
  030, [\href{http://arxiv.org/abs/1901.02253}{{\tt arXiv:1901.02253}}].

\bibitem{Mondini:2019vub}
R.~Mondini and C.~Williams, {\it {$ H\to b\overline{b}j $ at
  next-to-next-to-leading order accuracy}},  {\em JHEP} {\bf 06} (2019) 120,
  [\href{http://arxiv.org/abs/1904.08961}{{\tt arXiv:1904.08961}}].

\bibitem{0803.0883}
M.~Bahr et~al., {\it {Herwig++ Physics and Manual}},  {\em Eur. Phys. J. C}
  {\bf 58} (2008) 639--707, [\href{http://arxiv.org/abs/0803.0883}{{\tt
  arXiv:0803.0883}}].

\bibitem{Sjostrand:2014zea}
T.~Sj\"ostrand, S.~Ask, J.~R. Christiansen, R.~Corke, N.~Desai, P.~Ilten,
  S.~Mrenna, S.~Prestel, C.~O. Rasmussen, and P.~Z. Skands, {\it {An
  introduction to PYTHIA 8.2}},  {\em Comput. Phys. Commun.} {\bf 191} (2015)
  159--177, [\href{http://arxiv.org/abs/1410.3012}{{\tt arXiv:1410.3012}}].

\bibitem{0811.4622}
T.~Gleisberg, S.~Hoeche, F.~Krauss, M.~Schonherr, S.~Schumann, F.~Siegert, and
  J.~Winter, {\it {Event generation with SHERPA 1.1}},  {\em JHEP} {\bf 02}
  (2009) 007, [\href{http://arxiv.org/abs/0811.4622}{{\tt arXiv:0811.4622}}].

\bibitem{hep-ph/0204244}
S.~Frixione and B.~R. Webber, {\it {Matching NLO QCD computations and parton
  shower simulations}},  {\em JHEP} {\bf 06} (2002) 029,
  [\href{http://arxiv.org/abs/hep-ph/0204244}{{\tt hep-ph/0204244}}].

\bibitem{0709.2092}
S.~Frixione, P.~Nason, and C.~Oleari, {\it {Matching NLO QCD computations with
  Parton Shower simulations: the POWHEG method}},  {\em JHEP} {\bf 11} (2007)
  070, [\href{http://arxiv.org/abs/0709.2092}{{\tt arXiv:0709.2092}}].

\bibitem{1212.4504}
K.~Hamilton, P.~Nason, C.~Oleari, and G.~Zanderighi, {\it {Merging H/W/Z + 0
  and 1 jet at NLO with no merging scale: a path to parton shower + NNLO
  matching}},  {\em JHEP} {\bf 05} (2013) 082,
  [\href{http://arxiv.org/abs/1212.4504}{{\tt arXiv:1212.4504}}].

\bibitem{1309.0017}
K.~Hamilton, P.~Nason, E.~Re, and G.~Zanderighi, {\it {NNLOPS simulation of
  Higgs boson production}},  {\em JHEP} {\bf 10} (2013) 222,
  [\href{http://arxiv.org/abs/1309.0017}{{\tt arXiv:1309.0017}}].

\bibitem{1311.0286}
S.~Alioli, C.~W. Bauer, C.~Berggren, F.~J. Tackmann, J.~R. Walsh, and
  S.~Zuberi, {\it {Matching Fully Differential NNLO Calculations and Parton
  Showers}},  {\em JHEP} {\bf 06} (2014) 089,
  [\href{http://arxiv.org/abs/1311.0286}{{\tt arXiv:1311.0286}}].

\bibitem{1405.3607}
S.~H\"oche, Y.~Li, and S.~Prestel, {\it {Drell-Yan lepton pair production at
  NNLO QCD with parton showers}},  {\em Phys. Rev. D} {\bf 91} (2015), no.~7
  074015, [\href{http://arxiv.org/abs/1405.3607}{{\tt arXiv:1405.3607}}].

\bibitem{1407.3773}
S.~H\"oche, Y.~Li, and S.~Prestel, {\it {Higgs-boson production through gluon
  fusion at NNLO QCD with parton showers}},  {\em Phys. Rev. D} {\bf 90}
  (2014), no.~5 054011, [\href{http://arxiv.org/abs/1407.3773}{{\tt
  arXiv:1407.3773}}].

\bibitem{1407.2940}
A.~Karlberg, E.~Re, and G.~Zanderighi, {\it {NNLOPS accurate Drell-Yan
  production}},  {\em JHEP} {\bf 09} (2014) 134,
  [\href{http://arxiv.org/abs/1407.2940}{{\tt arXiv:1407.2940}}].

\bibitem{1508.01475}
S.~Alioli, C.~W. Bauer, C.~Berggren, F.~J. Tackmann, and J.~R. Walsh, {\it
  {Drell-Yan production at NNLL'+NNLO matched to parton showers}},  {\em Phys.
  Rev. D} {\bf 92} (2015), no.~9 094020,
  [\href{http://arxiv.org/abs/1508.01475}{{\tt arXiv:1508.01475}}].

\bibitem{1603.01620}
W.~Astill, W.~Bizon, E.~Re, and G.~Zanderighi, {\it {NNLOPS accurate associated
  HW production}},  {\em JHEP} {\bf 06} (2016) 154,
  [\href{http://arxiv.org/abs/1603.01620}{{\tt arXiv:1603.01620}}].

\bibitem{1805.09857}
E.~Re, M.~Wiesemann, and G.~Zanderighi, {\it {NNLOPS accurate predictions for
  $W^+W^-$ production}},  {\em JHEP} {\bf 12} (2018) 121,
  [\href{http://arxiv.org/abs/1805.09857}{{\tt arXiv:1805.09857}}].

\bibitem{1908.06987}
P.~F. Monni, P.~Nason, E.~Re, M.~Wiesemann, and G.~Zanderighi, {\it
  {MiNNLO$_{PS}$: a new method to match NNLO QCD to parton showers}},  {\em
  JHEP} {\bf 05} (2020) 143, [\href{http://arxiv.org/abs/1908.06987}{{\tt
  arXiv:1908.06987}}].

\bibitem{1909.02026}
S.~Alioli, A.~Broggio, S.~Kallweit, M.~A. Lim, and L.~Rottoli, {\it
  {Higgsstrahlung at NNLL'$+$NNLO matched to parton showers in GENEVA}},  {\em
  Phys. Rev. D} {\bf 100} (2019), no.~9 096016,
  [\href{http://arxiv.org/abs/1909.02026}{{\tt arXiv:1909.02026}}].

\bibitem{2006.04133}
P.~F. Monni, E.~Re, and M.~Wiesemann, {\it {MiNNLO$_{\text {PS}}$: optimizing
  $2\rightarrow 1$ hadronic processes}},  {\em Eur. Phys. J. C} {\bf 80}
  (2020), no.~11 1075, [\href{http://arxiv.org/abs/2006.04133}{{\tt
  arXiv:2006.04133}}].

\bibitem{2010.10478}
D.~Lombardi, M.~Wiesemann, and G.~Zanderighi, {\it {Advancing MiNNLO$_{\rm PS}$
  to diboson processes: $Z\gamma$ production at NNLO+PS}},
  \href{http://arxiv.org/abs/2010.10478}{{\tt arXiv:2010.10478}}.

\bibitem{2010.10498}
S.~Alioli, A.~Broggio, A.~Gavardi, S.~Kallweit, M.~A. Lim, R.~Nagar,
  D.~Napoletano, and L.~Rottoli, {\it {Precise predictions for photon pair
  production matched to parton showers in GENEVA}},
  \href{http://arxiv.org/abs/2010.10498}{{\tt arXiv:2010.10498}}.

\bibitem{1912.09982}
W.~Bizo\'n, E.~Re, and G.~Zanderighi, {\it {NNLOPS description of the $H \to
  b\overline{b} $ decay with MiNLO}},  {\em JHEP} {\bf 06} (2020) 006,
  [\href{http://arxiv.org/abs/1912.09982}{{\tt arXiv:1912.09982}}].

\bibitem{1206.3572}
K.~Hamilton, P.~Nason, and G.~Zanderighi, {\it {MINLO: Multi-Scale Improved
  NLO}},  {\em JHEP} {\bf 10} (2012) 155,
  [\href{http://arxiv.org/abs/1206.3572}{{\tt arXiv:1206.3572}}].

\bibitem{1002.2581}
S.~Alioli, P.~Nason, C.~Oleari, and E.~Re, {\it {A general framework for
  implementing NLO calculations in shower Monte Carlo programs: the POWHEG
  BOX}},  {\em JHEP} {\bf 06} (2010) 043,
  [\href{http://arxiv.org/abs/1002.2581}{{\tt arXiv:1002.2581}}].

\bibitem{2009.13533}
S.~Alioli, A.~Broggio, A.~Gavardi, S.~Kallweit, M.~A. Lim, R.~Nagar,
  D.~Napoletano, and L.~Rottoli, {\it {Resummed predictions for hadronic Higgs
  boson decays}},  \href{http://arxiv.org/abs/2009.13533}{{\tt
  arXiv:2009.13533}}.

\bibitem{1211.7049}
S.~Alioli, C.~W. Bauer, C.~J. Berggren, A.~Hornig, F.~J. Tackmann, C.~K.
  Vermilion, J.~R. Walsh, and S.~Zuberi, {\it {Combining Higher-Order
  Resummation with Multiple NLO Calculations and Parton Showers in GENEVA}},
  {\em JHEP} {\bf 09} (2013) 120, [\href{http://arxiv.org/abs/1211.7049}{{\tt
  arXiv:1211.7049}}].

\bibitem{hep-ph/9708255}
K.~Chetyrkin, B.~A. Kniehl, and M.~Steinhauser, {\it {Decoupling relations to O
  (alpha-s**3) and their connection to low-energy theorems}},  {\em Nucl. Phys.
  B} {\bf 510} (1998) 61--87, [\href{http://arxiv.org/abs/hep-ph/9708255}{{\tt
  hep-ph/9708255}}].

\bibitem{Kataev:1981gr}
A.~Kataev, N.~Krasnikov, and A.~Pivovarov, {\it {Two Loop Calculations for the
  Propagators of Gluonic Currents}},  {\em Nucl. Phys. B} {\bf 198} (1982)
  508--518, [\href{http://arxiv.org/abs/hep-ph/9612326}{{\tt hep-ph/9612326}}].
  [Erratum: Nucl.Phys.B 490, 505--507 (1997)].

\bibitem{Inami:1982xt}
T.~Inami, T.~Kubota, and Y.~Okada, {\it {Effective Gauge Theory and the Effect
  of Heavy Quarks in Higgs Boson Decays}},  {\em Z. Phys. C} {\bf 18} (1983)
  69--80.

\bibitem{Dawson:1990zj}
S.~Dawson, {\it {Radiative corrections to Higgs boson production}},  {\em Nucl.
  Phys. B} {\bf 359} (1991) 283--300.

\bibitem{Djouadi:1991tka}
A.~Djouadi, M.~Spira, and P.~Zerwas, {\it {Production of Higgs bosons in proton
  colliders: QCD corrections}},  {\em Phys. Lett. B} {\bf 264} (1991) 440--446.

\bibitem{Kataev:1993be}
A.~L. Kataev and V.~T. Kim, {\it {The Effects of the QCD corrections to Gamma
  (H0 ---\ensuremath{>} b anti-b)}},  {\em Mod. Phys. Lett. A} {\bf 9} (1994)
  1309--1326.

\bibitem{hep-ph/9405325}
L.~R. Surguladze, {\it {Quark mass effects in fermionic decays of the Higgs
  boson in O (alpha-s**2) perturbative QCD}},  {\em Phys. Lett. B} {\bf 341}
  (1994) 60--72, [\href{http://arxiv.org/abs/hep-ph/9405325}{{\tt
  hep-ph/9405325}}].

\bibitem{Spira:1995rr}
M.~Spira, A.~Djouadi, D.~Graudenz, and P.~Zerwas, {\it {Higgs boson production
  at the LHC}},  {\em Nucl. Phys. B} {\bf 453} (1995) 17--82,
  [\href{http://arxiv.org/abs/hep-ph/9504378}{{\tt hep-ph/9504378}}].

\bibitem{1408.5150}
J.~Gao and H.~X. Zhu, {\it {Electroweak prodution of top-quark pairs in e+e-
  annihilation at NNLO in QCD: the vector contributions}},  {\em Phys. Rev. D}
  {\bf 90} (2014), no.~11 114022, [\href{http://arxiv.org/abs/1408.5150}{{\tt
  arXiv:1408.5150}}].

\bibitem{1410.3165}
J.~Gao and H.~X. Zhu, {\it {Top Quark Forward-Backward Asymmetry in $e^+e^-$
  Annihilation at Next-to-Next-to-Leading Order in QCD}},  {\em Phys. Rev.
  Lett.} {\bf 113} (2014), no.~26 262001,
  [\href{http://arxiv.org/abs/1410.3165}{{\tt arXiv:1410.3165}}].

\bibitem{1408.5134}
A.~von Manteuffel, R.~M. Schabinger, and H.~X. Zhu, {\it {The two-loop soft
  function for heavy quark pair production at future linear colliders}},  {\em
  Phys. Rev. D} {\bf 92} (2015), no.~4 045034,
  [\href{http://arxiv.org/abs/1408.5134}{{\tt arXiv:1408.5134}}].

\bibitem{hep-ph/0508254}
W.~Bernreuther, R.~Bonciani, T.~Gehrmann, R.~Heinesch, P.~Mastrolia, and
  E.~Remiddi, {\it {Decays of scalar and pseudoscalar Higgs bosons into
  fermions: Two-loop QCD corrections to the Higgs-quark-antiquark amplitude}},
  {\em Phys. Rev. D} {\bf 72} (2005) 096002,
  [\href{http://arxiv.org/abs/hep-ph/0508254}{{\tt hep-ph/0508254}}].

\bibitem{1712.09889}
J.~Ablinger, A.~Behring, J.~Bl\"umlein, G.~Falcioni, A.~De~Freitas,
  P.~Marquard, N.~Rana, and C.~Schneider, {\it {Heavy quark form factors at two
  loops}},  {\em Phys. Rev. D} {\bf 97} (2018), no.~9 094022,
  [\href{http://arxiv.org/abs/1712.09889}{{\tt arXiv:1712.09889}}].

\bibitem{0803.0342}
T.~Becher and M.~D. Schwartz, {\it {A precise determination of $\alpha_s$ from
  LEP thrust data using effective field theory}},  {\em JHEP} {\bf 07} (2008)
  034, [\href{http://arxiv.org/abs/0803.0342}{{\tt arXiv:0803.0342}}].

\bibitem{Schwartz:2007ib}
M.~D. Schwartz, {\it {Resummation and NLO matching of event shapes with
  effective field theory}},  {\em Phys. Rev. D} {\bf 77} (2008) 014026,
  [\href{http://arxiv.org/abs/0709.2709}{{\tt arXiv:0709.2709}}].

\bibitem{Fleming:2007xt}
S.~Fleming, A.~H. Hoang, S.~Mantry, and I.~W. Stewart, {\it {Top Jets in the
  Peak Region: Factorization Analysis with NLL Resummation}},  {\em Phys. Rev.
  D} {\bf 77} (2008) 114003, [\href{http://arxiv.org/abs/0711.2079}{{\tt
  arXiv:0711.2079}}].

\bibitem{Kelley:2011ng}
R.~Kelley, M.~D. Schwartz, R.~M. Schabinger, and H.~X. Zhu, {\it {The two-loop
  hemisphere soft function}},  {\em Phys. Rev. D} {\bf 84} (2011) 045022,
  [\href{http://arxiv.org/abs/1105.3676}{{\tt arXiv:1105.3676}}].

\bibitem{Becher:2006qw}
T.~Becher and M.~Neubert, {\it {Toward a NNLO calculation of the anti-B
  ---\ensuremath{>} X(s) gamma decay rate with a cut on photon energy. II.
  Two-loop result for the jet function}},  {\em Phys. Lett. B} {\bf 637} (2006)
  251--259, [\href{http://arxiv.org/abs/hep-ph/0603140}{{\tt hep-ph/0603140}}].

\bibitem{Becher:2010pd}
T.~Becher and G.~Bell, {\it {The gluon jet function at two-loop order}},  {\em
  Phys. Lett. B} {\bf 695} (2011) 252--258,
  [\href{http://arxiv.org/abs/1008.1936}{{\tt arXiv:1008.1936}}].

\bibitem{Bruser:2018rad}
R.~Br\"user, Z.~L. Liu, and M.~Stahlhofen, {\it {Three-Loop Quark Jet
  Function}},  {\em Phys. Rev. Lett.} {\bf 121} (2018), no.~7 072003,
  [\href{http://arxiv.org/abs/1804.09722}{{\tt arXiv:1804.09722}}].

\bibitem{Banerjee:2018ozf}
P.~Banerjee, P.~K. Dhani, and V.~Ravindran, {\it {Gluon jet function at three
  loops in QCD}},  {\em Phys. Rev. D} {\bf 98} (2018), no.~9 094016,
  [\href{http://arxiv.org/abs/1805.02637}{{\tt arXiv:1805.02637}}].

\bibitem{Harlander:2003ai}
R.~V. Harlander and W.~B. Kilgore, {\it {Higgs boson production in bottom quark
  fusion at next-to-next-to leading order}},  {\em Phys. Rev. D} {\bf 68}
  (2003) 013001, [\href{http://arxiv.org/abs/hep-ph/0304035}{{\tt
  hep-ph/0304035}}].

\bibitem{Gehrmann:2005pd}
T.~Gehrmann, T.~Huber, and D.~Maitre, {\it {Two-loop quark and gluon
  form-factors in dimensional regularisation}},  {\em Phys. Lett. B} {\bf 622}
  (2005) 295--302, [\href{http://arxiv.org/abs/hep-ph/0507061}{{\tt
  hep-ph/0507061}}].

\bibitem{Moch:2005tm}
S.~Moch, J.~Vermaseren, and A.~Vogt, {\it {Three-loop results for quark and
  gluon form-factors}},  {\em Phys. Lett. B} {\bf 625} (2005) 245--252,
  [\href{http://arxiv.org/abs/hep-ph/0508055}{{\tt hep-ph/0508055}}].

\bibitem{Gehrmann:2010ue}
T.~Gehrmann, E.~Glover, T.~Huber, N.~Ikizlerli, and C.~Studerus, {\it
  {Calculation of the quark and gluon form factors to three loops in QCD}},
  {\em JHEP} {\bf 06} (2010) 094, [\href{http://arxiv.org/abs/1004.3653}{{\tt
  arXiv:1004.3653}}].

\bibitem{Gehrmann:2014vha}
T.~Gehrmann and D.~Kara, {\it {The $Hb\bar{b}$ form factor to three loops in
  QCD}},  {\em JHEP} {\bf 09} (2014) 174,
  [\href{http://arxiv.org/abs/1407.8114}{{\tt arXiv:1407.8114}}].

\bibitem{1210.2808}
J.~Gao, C.~S. Li, and H.~X. Zhu, {\it {Top Quark Decay at Next-to-Next-to
  Leading Order in QCD}},  {\em Phys. Rev. Lett.} {\bf 110} (2013), no.~4
  042001, [\href{http://arxiv.org/abs/1210.2808}{{\tt arXiv:1210.2808}}].

\bibitem{1501.07226}
V.~Del~Duca, C.~Duhr, G.~Somogyi, F.~Tramontano, and Z.~Tr\'ocs\'anyi, {\it
  {Higgs boson decay into b-quarks at NNLO accuracy}},  {\em JHEP} {\bf 04}
  (2015) 036, [\href{http://arxiv.org/abs/1501.07226}{{\tt arXiv:1501.07226}}].

\bibitem{hep-ph/9707448}
C.~R. Schmidt, {\it {H ---\ensuremath{>} g g g (g q anti-q) at two loops in the
  large M(t) limit}},  {\em Phys. Lett. B} {\bf 413} (1997) 391--395,
  [\href{http://arxiv.org/abs/hep-ph/9707448}{{\tt hep-ph/9707448}}].

\bibitem{1404.7096}
G.~Cullen et~al., {\it {G$\scriptsize{O}$S$\scriptsize{AM}$-2.0: a tool for
  automated one-loop calculations within the Standard Model and beyond}},  {\em
  Eur. Phys. J. C} {\bf 74} (2014), no.~8 3001,
  [\href{http://arxiv.org/abs/1404.7096}{{\tt arXiv:1404.7096}}].

\bibitem{1203.0291}
P.~Mastrolia, E.~Mirabella, and T.~Peraro, {\it {Integrand reduction of
  one-loop scattering amplitudes through Laurent series expansion}},  {\em
  JHEP} {\bf 06} (2012) 095, [\href{http://arxiv.org/abs/1203.0291}{{\tt
  arXiv:1203.0291}}]. [Erratum: JHEP 11, 128 (2012)].

\bibitem{1403.1229}
T.~Peraro, {\it {Ninja: Automated Integrand Reduction via Laurent Expansion for
  One-Loop Amplitudes}},  {\em Comput. Phys. Commun.} {\bf 185} (2014)
  2771--2797, [\href{http://arxiv.org/abs/1403.1229}{{\tt arXiv:1403.1229}}].

\bibitem{0903.4665}
A.~van Hameren, C.~Papadopoulos, and R.~Pittau, {\it {Automated one-loop
  calculations: A Proof of concept}},  {\em JHEP} {\bf 09} (2009) 106,
  [\href{http://arxiv.org/abs/0903.4665}{{\tt arXiv:0903.4665}}].

\bibitem{1007.4716}
A.~van Hameren, {\it {OneLOop: For the evaluation of one-loop scalar
  functions}},  {\em Comput. Phys. Commun.} {\bf 182} (2011) 2427--2438,
  [\href{http://arxiv.org/abs/1007.4716}{{\tt arXiv:1007.4716}}].

\bibitem{hep-ph/9512328}
S.~Frixione, Z.~Kunszt, and A.~Signer, {\it {Three jet cross-sections to
  next-to-leading order}},  {\em Nucl. Phys. B} {\bf 467} (1996) 399--442,
  [\href{http://arxiv.org/abs/hep-ph/9512328}{{\tt hep-ph/9512328}}].

\bibitem{hep-ph/0611247}
T.~Sjostrand, {\it {Monte Carlo Generators}},  in {\em {2006 European School of
  High-Energy Physics}}, pp.~51--74, 11, 2006.
\newblock \href{http://arxiv.org/abs/hep-ph/0611247}{{\tt hep-ph/0611247}}.

\bibitem{Tanabashi:2018oca}
{\bf Particle Data Group} Collaboration, M.~Tanabashi et~al., {\it {Review of
  Particle Physics}},  {\em Phys. Rev. D} {\bf 98} (2018), no.~3 030001.

\bibitem{hep-ph/0004189}
K.~Chetyrkin, J.~H. Kuhn, and M.~Steinhauser, {\it {RunDec: A Mathematica
  package for running and decoupling of the strong coupling and quark masses}},
   {\em Comput. Phys. Commun.} {\bf 133} (2000) 43--65,
  [\href{http://arxiv.org/abs/hep-ph/0004189}{{\tt hep-ph/0004189}}].

\bibitem{Skands:2014pea}
P.~Skands, S.~Carrazza, and J.~Rojo, {\it {Tuning PYTHIA 8.1: the Monash 2013
  Tune}},  {\em Eur. Phys. J. C} {\bf 74} (2014), no.~8 3024,
  [\href{http://arxiv.org/abs/1404.5630}{{\tt arXiv:1404.5630}}].

\bibitem{1003.0694}
A.~Buckley, J.~Butterworth, L.~Lonnblad, D.~Grellscheid, H.~Hoeth, J.~Monk,
  H.~Schulz, and F.~Siegert, {\it {Rivet user manual}},  {\em Comput. Phys.
  Commun.} {\bf 184} (2013) 2803--2819,
  [\href{http://arxiv.org/abs/1003.0694}{{\tt arXiv:1003.0694}}].

\bibitem{1707.01044}
F.~Herzog, B.~Ruijl, T.~Ueda, J.~Vermaseren, and A.~Vogt, {\it {On Higgs decays
  to hadrons and the R-ratio at N$^{4}$LO}},  {\em JHEP} {\bf 08} (2017) 113,
  [\href{http://arxiv.org/abs/1707.01044}{{\tt arXiv:1707.01044}}].

\bibitem{2011.10054}
K.~Hamilton, R.~Medves, G.~P. Salam, L.~Scyboz, and G.~Soyez, {\it {Colour and
  logarithmic accuracy in final-state parton showers}},
  \href{http://arxiv.org/abs/2011.10054}{{\tt arXiv:2011.10054}}.

\bibitem{2011.04773}
Z.~Nagy and D.~E. Soper, {\it {Summations of large logarithms by parton
  showers}},  \href{http://arxiv.org/abs/2011.04773}{{\tt arXiv:2011.04773}}.

\bibitem{2011.04777}
Z.~Nagy and D.~E. Soper, {\it {Summations by parton showers of large logarithms
  in electron-positron annihilation}},
  \href{http://arxiv.org/abs/2011.04777}{{\tt arXiv:2011.04777}}.

\bibitem{2011.15087}
J.~Holguin, J.~R. Forshaw, and S.~Pl\"atzer, {\it {Improvements on dipole
  shower colour}},  \href{http://arxiv.org/abs/2011.15087}{{\tt
  arXiv:2011.15087}}.

\bibitem{2002.11114}
M.~Dasgupta, F.~A. Dreyer, K.~Hamilton, P.~F. Monni, G.~P. Salam, and G.~Soyez,
  {\it {Parton showers beyond leading logarithmic accuracy}},  {\em Phys. Rev.
  Lett.} {\bf 125} (2020), no.~5 052002,
  [\href{http://arxiv.org/abs/2002.11114}{{\tt arXiv:2002.11114}}].

\bibitem{2003.06400}
J.~R. Forshaw, J.~Holguin, and S.~Pl\"atzer, {\it {Building a consistent parton
  shower}},  {\em JHEP} {\bf 09} (2020) 014,
  [\href{http://arxiv.org/abs/2003.06400}{{\tt arXiv:2003.06400}}].

\bibitem{2003.00702}
H.~Brooks, C.~T. Preuss, and P.~Skands, {\it {Sector Showers for Hadron
  Collisions}},  {\em JHEP} {\bf 07} (2020) 032,
  [\href{http://arxiv.org/abs/2003.00702}{{\tt arXiv:2003.00702}}].

\bibitem{Korchemsky:1987wg}
G.~Korchemsky and A.~Radyushkin, {\it {Renormalization of the Wilson Loops
  Beyond the Leading Order}},  {\em Nucl. Phys. B} {\bf 283} (1987) 342--364.

\bibitem{0903.2561}
N.~Kidonakis, {\it {Two-loop soft anomalous dimensions and NNLL resummation for
  heavy quark production}},  {\em Phys. Rev. Lett.} {\bf 102} (2009) 232003,
  [\href{http://arxiv.org/abs/0903.2561}{{\tt arXiv:0903.2561}}].

\bibitem{0904.1021}
T.~Becher and M.~Neubert, {\it {Infrared singularities of QCD amplitudes with
  massive partons}},  {\em Phys. Rev. D} {\bf 79} (2009) 125004,
  [\href{http://arxiv.org/abs/0904.1021}{{\tt arXiv:0904.1021}}]. [Erratum:
  Phys.Rev.D 80, 109901 (2009)].

\bibitem{Becher:2011pf}
T.~Becher, G.~Bell, and M.~Neubert, {\it {Factorization and Resummation for Jet
  Broadening}},  {\em Phys. Lett.} {\bf B704} (2011) 276--283,
  [\href{http://arxiv.org/abs/1104.4108}{{\tt arXiv:1104.4108}}].

\end{thebibliography}\endgroup
\bibliographystyle{jhep}

\end{document}